\documentclass[12pt]{article}

\usepackage{fullpage}
\usepackage{latexsym}
\usepackage{graphicx}
\usepackage[intlimits]{amsmath}
\usepackage{amsthm}
\usepackage{natbib}
\usepackage{amsfonts}
\usepackage{ulem}
\usepackage{booktabs}
\usepackage{placeins}
\usepackage{hyperref}
\usepackage{url}
\usepackage{setspace}
\usepackage{subcaption}
\usepackage{multirow, makecell}
\usepackage{color, colortbl}

\newcommand{\Exp}{\ensuremath{\text{Exp}}}
\newcommand{\Ngam}{\ensuremath{N_{\boldsymbol{\gamma}}}}

\definecolor{Gray}{gray}{0.9}

\begin{document}

\begin{center}
   {\LARGE Estimating effective population size changes from preferentially sampled genetic sequences}
   \\\ \\
  { Michael D. Karcher\textsuperscript{1},
  Marc A. Suchard\textsuperscript{2,3,4},
  Gytis Dudas\textsuperscript{5,6}, \\
  Vladimir N. Minin\textsuperscript{7,*}}\\ \ \\
  {\footnotesize \textsuperscript{1}Department of Statistics, University of Washington, Seattle\\
   \textsuperscript{2}Department of Human Genetics, David Geffen School of Medicine at UCLA,\\
   University of California, Los Angeles\\
   \textsuperscript{3}Department of Biomathematics, David Geffen School of Medicine at UCLA,\\
   University of California,  Los Angeles\\
      \textsuperscript{4}Department of Biostatistics, UCLA Fielding School of Public Health,\\
      University of California, Los Angeles\\
      \textsuperscript{5}Vaccine and Infectious Disease Division, Fred Hutchinson Cancer Research Center\\
      $^6$Gothenburg Global Biodiversity Centre (GGBC), Gothenburg, Sweden\\
     \textsuperscript{7}Department of Statistics, University of California, Irvine\\
     \textsuperscript{*}corresponding author: \url{vminin@uci.edu}}
\end{center}

\begin{abstract}
Coalescent theory combined with statistical modeling allows us
to estimate effective population size fluctuations from molecular sequences
of individuals sampled from a population of interest.
When sequences are sampled serially through time and
the distribution of the sampling times depends on the effective population size,
explicit statistical modeling of sampling times improves population size estimation.
Previous work assumed that the genealogy relating sampled sequences is known and
modeled sampling times as an inhomogeneous Poisson process with
log-intensity equal to a linear function of the log-transformed effective population size.
We improve this approach in two ways.
First, we extend the method to allow for joint Bayesian estimation of the genealogy,
effective population size trajectory, and other model parameters.
Next, we improve the sampling time model by incorporating additional sources of information
in the form of time-varying covariates.
We validate our new modeling framework using a simulation study and apply our new methodology to analyses of population dynamics of seasonal influenza and to the recent Ebola virus outbreak in West Africa.
\end{abstract}


\section{Introduction}

Phylodynamic inference---the study and estimation of population dynamics
from genetic sequences---relies upon data sampled in a timeframe compatible with
the evolutionary dynamics under question \citep{drummond2003measurably}.
One important class of phylodynamic methods seeks to estimate magnitudes and changes
in a measure of genetic diversity called the \textit{effective population size},
often considered proportional to the census population size \citep{wakeley2009extensions}
or number of infections in epidemiological contexts \citep{frost2010viral}.
One subtle and often ignored complication of phylodynamic inference
occurs when there is a probabilistic dependence between the effective population trajectory
and the temporal frequency of collecting data samples,
such as in case of sampling infectious disease agent genetic sequences with increasing urgency
and intensity during a rising epidemic.
This issue of \textit{preferential sampling} was studied in depth by \citet{karcher2015quantifying} in
the limited context of a known, fixed genealogy reconstructed from the genetic data,
revealing that sampling protocols that (implicitly) depend on effective population size
cause model misspecification bias in models that do not account for
the possibility of preferential sampling.
Here, we extend the work of \citet{karcher2015quantifying} and
develop a Bayesian framework for accounting for preferential sampling during
effective population size estimation directly from sequence data rather than
from a fixed genealogy.
We also propose a more flexible model for sequence sampling times that
allows for inclusion of arbitrary time-dependent covariates and their interactions with the effective population size.

Methods for estimating effective population size from
genealogical data and genetic sequence data have evolved from
the earliest low dimensional parametric methods,
such as constant population size \citep{griffiths1994sampling} and
exponential growth models \citep{griffiths1994sampling, drummond2002estimating},
to more flexible, nonparametric or highly parametric methods
based on change-point models and Gaussian process smoothing
\citep{drummond2005bayesian, Heled2008, skyride, palacios2013gaussian, palacios2012INLA, skygrid, Gill2016}.
Most coalescent-based methods condition on the times of sequence sampling,
rather than include these times into the model,
leaving open the possibility of model misspecification if preferential sampling over time is in play.
\citet{volz2014sampling} and \citet{karcher2015quantifying} introduced
coalescent models that include sampling times as random variables,
whose distribution is allowed to depend on the effective population size.
In particular, \citet{karcher2015quantifying} propose a method that
models sampling times as an inhomogeneous Poisson process
with log-intensity equal to an affine transformation of the log-transformed effective population size.
In the presence of preferential sampling,
this sampling-aware model demonstrates improved accuracy and precision
compared to standard coalescent models
due to eliminating an element of model misspecification and
incorporating an additional source of information to estimate the effective population trajectory.

The main limitations of the approach of \citet{karcher2015quantifying} are
a reliance on a fixed, known genealogy
and lack of flexibility in the preferential sampling time model that
currently does not allow the relationship between
effective population size and sampling intensity to change over time.
We address the issue of fixed-tree inference by implementing a preferential sampling time model
in the popular phylodynamic Markov chain Monte Carlo (MCMC) software package BEAST \citep{BEAST}.
This allows us to perform inference directly from genetic sequence data,
appropriately accounting for genealogical uncertainty,
using a wide selection of molecular sequence evolution models and
well tested phylogenetic MCMC transition kernels.
Additionally, we implement a tuning parameter free elliptical slice sampling transition kernel
\citep{murray2010elliptical} for high dimensional effective population size trajectory parameters,
which allows us to update these parameters efficiently.

We also address the issue of an inflexible preferential sampling time model
by incorporating time-varying covariates into the model.
We model the sampling times as an inhomogeneous Poisson process
with log-intensity equal to a linear combination of the log-effective population size and any number of functions of time.
These functions can include time varying \textit{covariates} and products of covariates and the log-effective population size,
referred to as \textit{interaction covariates}.
The addition of covariates into the sampling time model
allows for incorporating additional sources of information into the relationship between
effective population size and sampling intensity.
One example of time-varying covariates includes
an exponential growth function to account for
a continuous decrease in sequencing costs that
results in increased intensity of genetic data collection over time.
In the context of endemic infectious disease surveillance,
it is likely important to account for seasonality when
modeling changes in genetic data sampling intensity,
motivating inclusion of periodic functions as time varying covariates in the preferential sampling model.

We validate our methods first by simulating genealogies and sequence data
and confirming that our methods successfully reconstruct the
true effective population trajectories and true model parameters.
We briefly simulate data in a fixed-tree context to demonstrate the fundamentals of
incorporating covariates into the sampling time model and what bias is introduced by
model misspecifications.
We proceed to simulate genetic sequence data and demonstrate that our model
successfully functions when we estimate effective population size trajectory and
other parameters directly from sequence data.
We also use simulations to test a combination of the two extensions of the preferential sampling model and
work with covariates while sampling over genealogies during the MCMC.
Finally, we use our method to analyze two real-world epidemiological datasets.
We analyze a USA/Canada regional influenza dataset \citep{zinder2014seasonality}
to determine if exponential growth of genetic sequencing or
seasonal changes in sampling intensity are important to adjust for during
effective population size reconstruction.
We also analyze data from the recent Ebola outbreak in Western Africa
to determine if preferential sampling has taken place and whether
time-varying covariates or interaction covariates improve the phylodynamic inference.

\section{Methods}
\label{sec:methods}

\subsection{Sequence Data and Substitution Model}
\label{sec:seq_data}

Consider an \textit{alignment} $\mathbf{y} = \{y_{ij}\}$, $i=1,\dots,n$, $j=1,\dots,l$,
of $n$  genetic sequences across $l$ sites,
collected from a well-mixed population at \textit{sampling times}
\[{\mathbf{s} = \{s_i\}_{i=1}^{n}, \: \allowbreak s_1 \geq \ldots \geq s_n = 0.}\]
The following example
shows an alignment of $n=5$ samples across $l=10$ sites,
sampled at distinct times between time 7 and time 0---with
time understood to be time \textit{before} the latest sample:
\begin{alignat*}{2}
\mathbf{y}_1 = \: &ACATGAGCTT, \; &&s_1 = 7 \\
\mathbf{y}_2 = \: &ACTTGACCTG, \; &&s_2 = 4 \\
\mathbf{y}_3 = \: &TCTTGACCTT, \; &&s_3 = 2 \\
\mathbf{y}_4 = \: &AAATCTGCGT, \; &&s_4 = 1 \\
\mathbf{y}_5 = \: &AGATGTGCAT, \; &&s_5 = 0. \\
\end{alignat*}

\begin{figure}[tbh]
	\centering
	\includegraphics[width=0.9\textwidth]{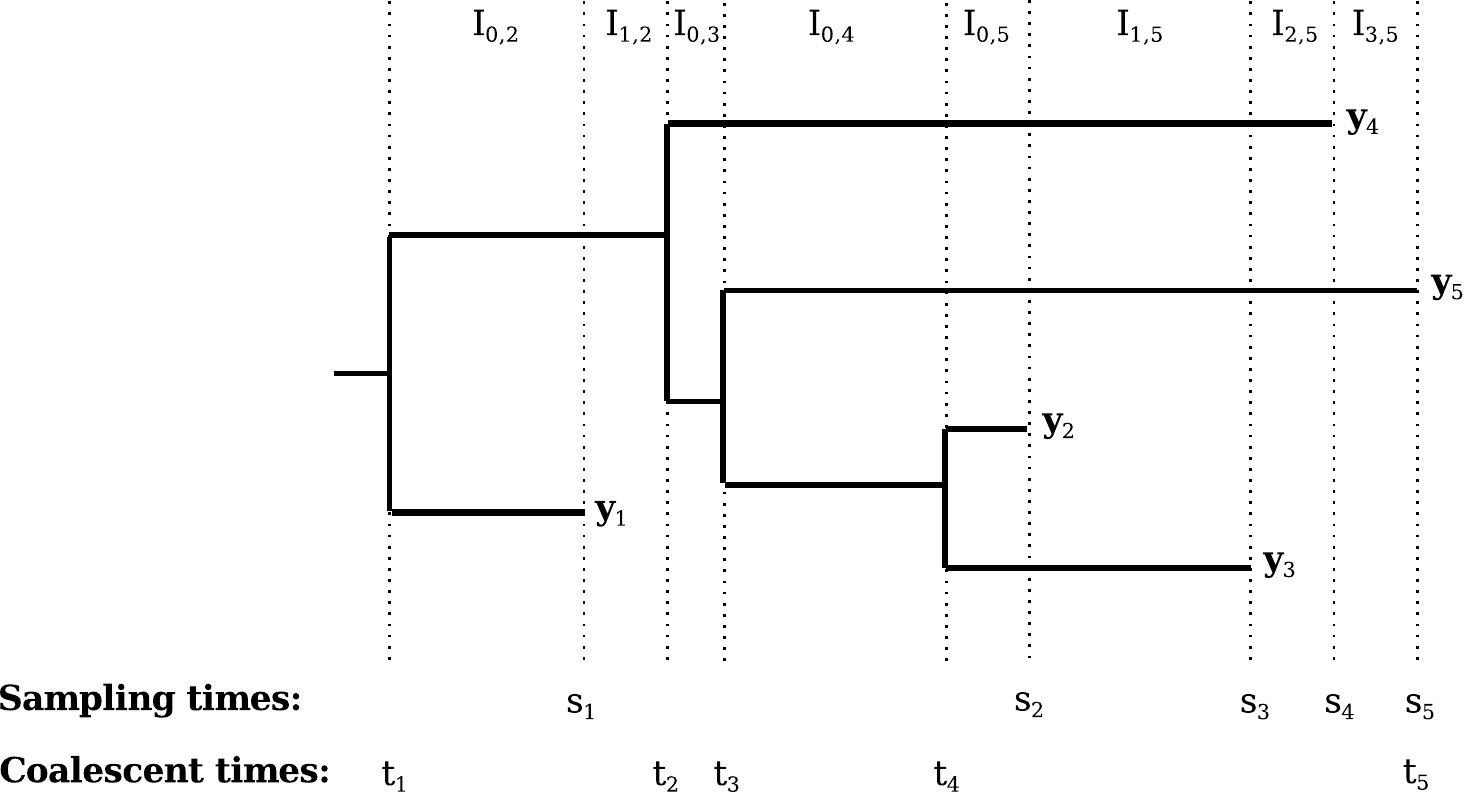}
	\caption{{\bf Illustration of an example heterochronous genealogy with $n = 5$ lineages.}
Sampling times $s_1,\ldots,s_5$ and coalescent times $t_1,\ldots,t_5$ are marked below the genealogy,
and sequence data $\mathbf{y}_1,\ldots,\mathbf{y}_5$ are marked at their corresponding tips.}
	\label{fig:Genealogy}
\end{figure}

All of the individual sequences share a common ancestry,
which can be represented by a bifurcating tree called a \textit{genealogy}---illustrated
in Figure \ref{fig:Genealogy}.

We assume that sequence data $\mathbf{y}$ are generated by a continuous time Markov chain
(CTMC) \textit{substitution model} that models the evolution of the genetic sequence
along the genealogy $\mathbf{g}$.
According to this model, alignment sites are independent and identically distributed,
with a transition matrix $\boldsymbol{\theta}$ controlling the CTMC substitution rates
between the different nucleotide bases.
Some relaxation of these assumptions is possible \citep{Shapiro2005}.
Different substitution models are then defined by different parameterizations of $\boldsymbol{\theta}$
\citep{hein2004gene}.
It is simple to simulate from these models,
and we can efficiently compute the probability of the observed sequence data $\mathbf{y}$,
\[
\text{Pr}(\mathbf{y} \mid \mathbf{g}, \boldsymbol{\theta}),
\]
using Felsenstein's pruning algorithm \citep{felsenstein1973maximum, felsenstein1981evolutionary}.

\subsection{The Coalescent}
\label{sec:coalescent}

Recall that we assume that the $n$ sampled sequences share a common ancestry,
which can be represented by a bifurcating tree called a \textit{genealogy}---illustrated
in Figure \ref{fig:Genealogy}.
The branching events of the tree
$\mathbf{g} = \{t_i\}_{i=1}^{n-1}, \allowbreak {t_1 > \ldots > t_{n-1}}$
(with $t$ greater the farther \textit{back} in time an event occurs)
are called \textit{coalescent events}.
The times associated with the tips of the tree $\mathbf{s} = \{s_i\}_{i=1}^{n}$,
$s_1 \geq \ldots \geq s_n$
are called \textit{sampling times} or \textit{sampling events}.
If all of the sampling events are simultaneous,
the sampling is called \textit{isochronous}.
Assuming that the population evolves according to
the Wright-Fisher model of genetic drift and that the size of the population is not changing,
\citet{coalescent} derived a probability density for an isochronous genealogy,
where the population size plays the role of a parameter of this density.
Since the Wright-Fisher model is a simplified representation of the evolutionary process,
the above parameter is called the \textit{effective population size}, $N_e$.
Later extensions to the coalescent model incorporated
variable effective population size $N_e(t)$ \citep{griffiths1994sampling} and
the ability to evaluate densities of  genealogies with \textit{heterochronously} sampled tips---genealogies with
non-simultaneous sampling times \citep{joseph_coalescent_1999}.

Given sampling times $\mathbf{s}$ and effective population size trajectory $N_e(t)$,
we would like to define the probability density for a particular genealogy $\mathbf{g}$.
We use the term \textit{active lineages}, $n(t)$,
to refer to the difference between the number of samples taken and
the number of coalescent events occurred between times $0$ and $t$.
To illustrate, in Figure \ref{fig:Genealogy},
$n(t)$ can be seen as the number of horizontal lines
that a vertical line at time $t$ will cross.
Suppose we partition the interval $(s_n, t_1)$,
from the most recent sampling event to the
\textit{time to most recent common ancestor} (TMRCA),
into intervals $I_{i,k}$ with constant numbers of active lineages.
Let $\lambda_c(t) = \binom{n(t)}{2}/N_e(t)$.
Then the coalescent density evaluated at genealogy $\mathbf{g}$ is
\begin{equation}
\Pr(\mathbf{g} \mid N_e(t), \mathbf{s}) \propto
         \prod_{k=2}^n {\left[\lambda_c(t_{k-1}) \exp\left(-\int_{I_{i,k}} {\lambda_c(t)dt}\right)\right]}.
         \label{eq:coalescent}
\end{equation}

\subsection{Population Size Prior}
\label{sec:pop_size}

Note that without further assumptions the effective population size trajectory
function $N_e(t)$ is infinite-dimensional,
so inference about $N_e(t)$ without some manner of constraint is intractable.
A number of approaches, reviewed in the Introduction,
have been suggested to address this fact.
Here, we take a regular grid approach that was used before in multiple studies
\citep{palacios2012INLA, skygrid, Gill2016, karcher2015quantifying}.
To review, we approximate $N_e(t)$ with a piecewise constant function,
$\Ngam(t) = \exp[\boldsymbol{\gamma}(t)]$,
where $\boldsymbol{\gamma}(t) = \sum_{i=1}^p{\gamma_i 1_{\{t \in J_i\}}}$ and
$J_1, \ldots, J_p$ are consecutive time intervals of equal length.
In contexts where the genealogy is known,
we choose intervals that perfectly cover the interval between the TMRCA and the latest sample.
However, in contexts where the genealogy is estimated from sequence data,
the TMRCA is not necessarily fixed.
To address this, we choose equal intervals that extend to a fixed point in time
and append an additional interval that extends from that point infinitely back in time.
This allows us to estimate the effective population trajectory with user-defined resolution over
a window that extends back in time as far as the user chooses.
The choice of the end point of the grid is up to the user,
but it is advisable to choose a point that is farther back in time than an \textit{a priori}
estimate of the TMRCA in order to extend the high-resolution grid to cover
the entire true genealogy.

The population size trajectory $\Ngam(t)$ is parameterized by a potentially
high dimensional vector $\boldsymbol{\gamma} = (\gamma_1, \ldots, \gamma_p)$.
We assume that \textit{a priori} $\boldsymbol{\gamma}$ follows
a first order Gaussian random walk prior with precision hyperparameter $\kappa$:
$\gamma_i  \mid \gamma_{i-1} \sim \mathcal{N}(\gamma_{i-1}, 1/\kappa)$
or, equivalently, that $\gamma_i - \gamma_{i-1} \sim \mathcal{N}(0, 1/\kappa)$,
for $i = 2,\dots, p$.
We use a Gaussian prior on the first element: $\gamma_1 \sim \mathcal{N}(0, \sigma_p^2)$.
Finally, we assign a $\text{Gamma}(\alpha, \beta)$ hyperprior to $\kappa$.

\subsection{Preferential Sampling Model with Covariates}
\label{sec:pref_samp}

\citet{karcher2015quantifying} model times at which sequences are collected as
a Poisson point process with intensity $\lambda_s(t)$ equal to
a log-linear function of the log effective population size.
Although it is realistic to assume that the larger the population,
the more members of the population gets sequenced,
other factors may influence the distribution of sequence sampling times.
For instance, decreasing sequencing costs may result in
increasing sequence sampling intensity even if the population size remains constant.
We propose an extension to the sampling model that allows for the incorporation
of time-varying covariates as additional sources of information.
Suppose we have one or more real-valued functions, $\mathcal{F} = \{f_2(t), \ldots, f_m(t)\}$.
We let
\begin{equation}
\log \lambda_s(t; \mathcal{F}) = \beta_0 + \beta_1 \gamma(t) + \beta_2 f_2(t) + \ldots + \beta_m f_m(t) + \left[\delta_2 f_2(t) + \ldots + \delta_m f_m(t)\right] \gamma(t),
    \label{eq:samplingintensity}
\end{equation}
where we may set any or all of the $\beta_2, \dots, \beta_m$ or
$\delta_2, \dots, \delta_m$ to zero if we want to avoid modeling effects
of certain covariates or their interactions with the log-population size.
Notice that we reserve $f_1(t)$ for $\gamma(t) = \log[N_e(t)]$,
which is the covariate that is always present in our model.
We also point out that even though Equation~\eqref{eq:samplingintensity} is written in continuous time,
in practice we assume that both the sampling intensity $\lambda_s(t)$ and our time varying covariates are piecewise constant,
with changes occurring at the grid points specified in Subsection \ref{sec:pop_size}.
We assign independent $\mathcal{N}(0, \sigma_s^2)$ priors for
all components of the preferential sampling model parameter vector
$\boldsymbol{\beta} = (\beta_0, \beta_1,\dots, \beta_m, \delta_2, \dots, \delta_m)$.

\subsection{Posterior Approximation with MCMC}
\begin{figure}[htb]
	\centering
	\includegraphics[width=0.6\textwidth]{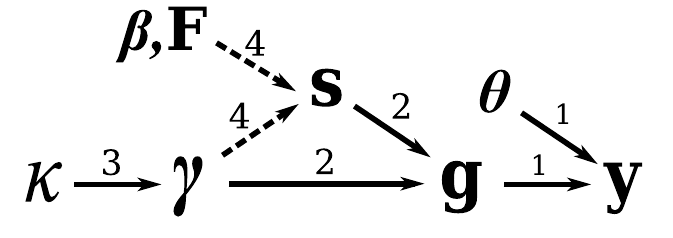}
	\caption{{\bf Dependency graph for the phylodynamic model parameters and data.}
	Dependencies labeled 1 are explored in section \ref{sec:seq_data},
	those labeled 2 are explored in section \ref{sec:coalescent},
	those labeled 3 are explored in section \ref{sec:pop_size},
	and those labeled 4 are explored in section \ref{sec:pref_samp}.
The dashed lines between $\boldsymbol{\gamma},\boldsymbol{\beta},\mathcal{F}$ and $\mathbf{s}$
represent preferential sampling.}
	\label{fig:DependencyGraph}
\end{figure}

Having specified all parts of our data generating model,
we are now ready to define the posterior distribution of all unknown variables of interest:
\begin{equation}
\begin{split}
\Pr(\mathbf{g}, \boldsymbol{\gamma}, \kappa, \boldsymbol{\beta}, \boldsymbol{\theta} \mid \mathbf{y}, \mathbf{s}, \mathcal{F}) \propto
    & \Pr(\mathbf{y} \mid \mathbf{g}, \boldsymbol{\theta})
    \Pr(\mathbf{g} \mid \boldsymbol{\gamma}, \mathbf{s})
    \Pr(\mathbf{s} \mid \boldsymbol{\gamma}, \boldsymbol{\beta}, \mathcal{F})
    \Pr(\boldsymbol{\gamma} \mid \kappa) \\
    &\times \Pr (\kappa) \Pr(\boldsymbol{\beta}) \Pr (\boldsymbol{\theta}),
\end{split}
    \label{eq:mainposteriorsamp}
\end{equation}
where all probabilities and probability densities on the righthand side of equation \eqref{eq:mainposteriorsamp} are defined in the previous subsections. Figure~\ref{fig:DependencyGraph} illustrates conditional dependencies of model parameters and data in a graph form.

When the distribution of sampling times does not depend on the effective population size trajectory
(in our model, this happens when $\beta_1 = 0$ and $\delta_2 = \cdots = \delta_m = 0$),
the posterior takes the following form:
\begin{equation*}
\begin{split}
\Pr(\mathbf{g}, \boldsymbol{\gamma}, \kappa, \boldsymbol{\theta},  \boldsymbol{\beta} \mid \mathbf{y}, \mathbf{s}, \mathcal{F}) &\propto
    \underbrace{\Pr(\mathbf{y} \mid \mathbf{g}, \boldsymbol{\theta})\Pr(\mathbf{g} \mid \boldsymbol{\gamma}, \mathbf{s})\Pr(\boldsymbol{\gamma} | \kappa) \Pr (\kappa) \Pr (\boldsymbol{\theta})}_{\propto \Pr(\mathbf{g}, \boldsymbol{\gamma}, \kappa, \boldsymbol{\theta} \mid \mathbf{y}, \mathbf{s})}\\
    &\times \underbrace{\Pr(\mathbf{s} \mid \boldsymbol{\beta}, \mathcal{F})
    \Pr(\boldsymbol{\beta})}_{\propto \Pr(\boldsymbol{\beta} \mid \mathbf{s}, \mathcal{F})}.
\end{split}
\end{equation*}
The factorization above demonstrates that when $\boldsymbol{\gamma}$
is absent from the $\text{Pr}(\mathbf{s} \mid \cdot)$ term,
joint and separate estimations of effective population size parameters
$\boldsymbol{\gamma}$ and preferential sampling model parameters
$\boldsymbol{\theta}$ will yield identical results.
Moreover, in this case estimation of sampling model parameters can be
dropped from the analysis entirely,
since typically these parameters would be considered nuisance.
If we drop preferential sampling,
our model specifications reduces to the Bayesian skygrid model of \citet{skygrid},
with the corresponding posterior:
\begin{equation}
\Pr(\mathbf{g}, \boldsymbol{\gamma}, \kappa, \boldsymbol{\theta} \mid \mathbf{y}, \mathbf{s})
 \propto
\Pr(\mathbf{y} \mid \mathbf{g}, \boldsymbol{\theta})\Pr(\mathbf{g} \mid \boldsymbol{\gamma}, \mathbf{s})\Pr(\boldsymbol{\gamma} | \kappa) \Pr (\kappa) \Pr (\boldsymbol{\theta}).
\label{eq:noprefsampling_posterior}
\end{equation}

We approximate posteriors \eqref{eq:mainposteriorsamp} and \eqref{eq:noprefsampling_posterior} by devising MCMC algorithms,
implemented in the software package BEAST
\citep{BEAST}, that target these distributions.
We update model parameters in blocks --- 1) genealogy $\mathbf{g}$,
2) substitution parameters $\boldsymbol{\theta}$,
3) population size parameters $\boldsymbol{\gamma}$,
4) random walk prior precision $\kappa$,
5) preferential sampling model parameters $\boldsymbol{\beta}$
--- keeping parameters outside of the block fixed.
We update the genealogy and substitution model parameters via
the default BEAST Markov kernels.
We update the log effective population latent field $\boldsymbol{\gamma}$
via an elliptical slice sampler (ESS) operator \citep{murray2010elliptical, lan2015efficient},
which takes advantage of the Gaussian prior distribution of the latent field to perform efficient updates.
Informally, it does this by sampling a set of parameter values from the prior and
iteratively moving the values closer to the current values via elliptical interpolation if
the coalescent likelihood falls below a random, but small,
neighborhood of the current likelihood.
Because the stepwise differences of the log effective population size trajectory,
$\Delta \boldsymbol{\gamma}$, are modeled as independent Gaussians with precision $\kappa$,
and because we give $\kappa$ a Gamma($\alpha$, $\beta$) hyperprior,
we update $\kappa$ using a Normal-Gamma Gibbs update kernel with full conditional
\[
\kappa \mid \Delta \boldsymbol{\gamma} \sim
       \text{Gamma}\left[\alpha + \frac{p}{2}, \beta + \frac{1}{2} \sum_{i=2}^p{ \left( \gamma_i - \gamma_{i-1} \right)^2}\right],
\]
where $p$ is the number of parameters in the latent field.
For our sampling conditional model with posterior \eqref{eq:noprefsampling_posterior},
we finish here and refer to the method as \textit{ESS/BEAST},
abbreviated when appropriate as \textit{ESS}.
For our sampling-aware model with the posterior \eqref{eq:mainposteriorsamp},
we update components of the preferential sampling model parameter vector
$\boldsymbol{\beta}$ with univariate Gaussian random walk Metropolis-Hastings kernels.
We refer to the method as \textit{SampESS/BEAST},
abbreviated when appropriate as \textit{SampESS}.

\section{Implementation}
We implemented INLA-based, fixed-genealogy BNPR-PS method with simple covariates
in R package \texttt{phylodyn} (\url{https://github.com/mdkarcher/phylodyn}).
The package has also MCMC functionality that can handle inference
from a fixed genealogy with simple and interaction sampling model covariates.
See \texttt{phylodyn} vignettes for more details.
MCMC for direct inference from sequence data is available in the development branch
of software package \texttt{BEAST} (\url{https://github.com/beast-dev/beast-mcmc}).
We provide examples of how to specify our preferential sampling models
in \texttt{BEAST} xml files at \url{https://github.com/mdkarcher/BEAST-XML}.

\section{Results}

\subsection{Simulation Study}

\subsubsection{Inference Assuming Fixed Genealogy}

In Section~\ref{sec:pref_samp}, we proposed an extended sampling time model
that incorporated time-varying covariates.
We perform a simulation study to confirm the ability of our method to recover
the true effective population trajectory and model coefficients with covariates affecting
the sampling intensity.
We begin here with fixed genealogies and move on to
direct inference from sequence data in the next section.

We start with the inhomogeneous Poisson process sampling model
with log-intensity as in Equation~\ref{eq:samplingintensity}.
If we restrict all $\beta$s and $\delta$s to be zero aside from $\beta_0$,
the model collapses to homogenous Poisson process sampling
(equivalently, uniformly sampling a Poisson number of points across the sampling interval).
If we allow $\beta_1$ to be nonzero, the model becomes the sampling-aware model
of \citet{karcher2015quantifying}.
If we allow additional $\beta$s,
each corresponding to a fixed function of time,
to be nonzero (but not $\delta$s)
we say that the model includes \textit{simple} or \textit{ordinary covariates}.

\begin{figure}
	\centering
	\includegraphics[width=1.0\textwidth]{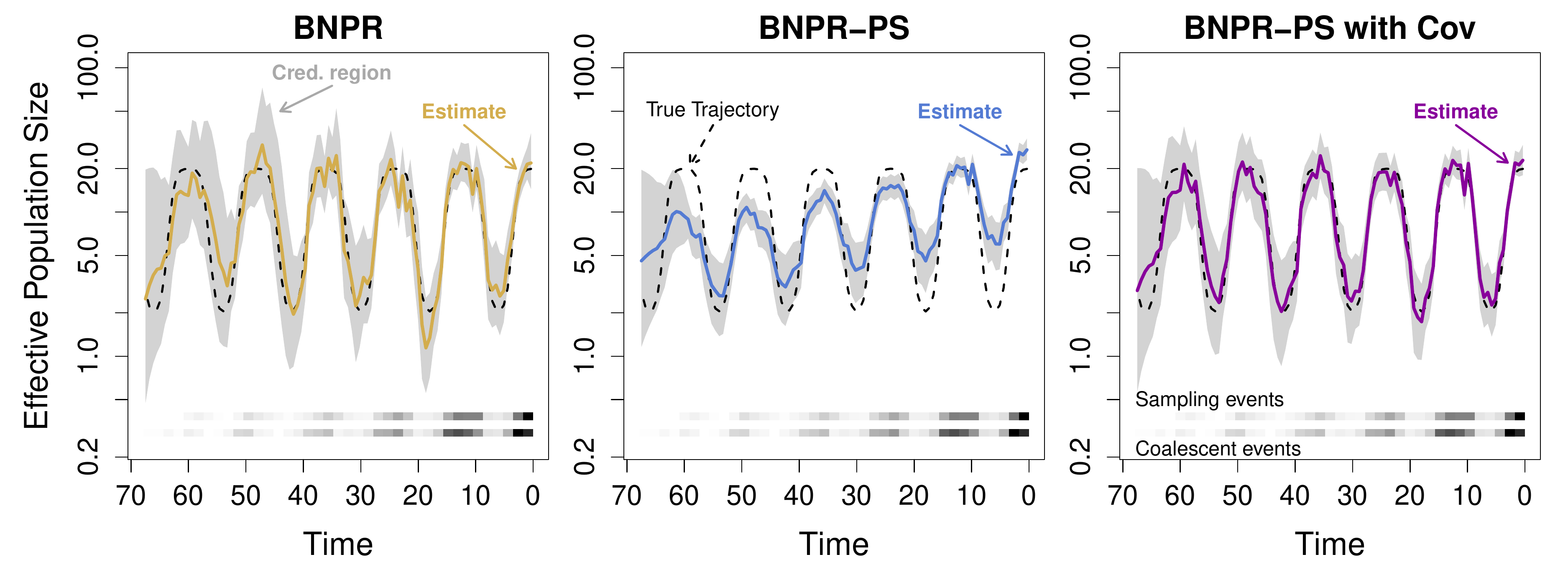}
	\caption{{\bf Effective population size reconstruction for BNPR, BNPR-PS, and BNPR-PS with simple covariates.}
		The dotted black line represents the true effective population trajectory.
		The solid colored line represents the marginal posterior median effective population trajectory
		inferred by BNPR (yellow), BNPR-PS (blue), and BNPR-PS with simple covariates (purple),
		and the gray region represents the corresponding pointwise 95\% credible intervals
		for the effective population trajectory.
		The log sampling intensity was $1.557 + \gamma(t) - 0.025t$.}
	\label{fig:SimpleCovariatesComparison}
\end{figure}

For computational efficiency in this simulation study,
we build upon the methods of \citet{karcher2015quantifying},
including Bayesian Nonparametric Population Reconstruction (BNPR)
which uses integrated nested Laplace approximation (INLA) to efficiently approximate
the marginal posterior for fixed-genealogy data,
and Bayesian Nonparametric Population Reconstruction with Preferential Sampling (BNPR-PS)
which does the same but includes our sampling time model (without covariates).
We incorporate our extended sampling time model into BNPR-PS,
but due to constraints in the INLA R package, upon which BNPR-PS relies,
we can only include simple covariates.

Because our sampling time model is an inhomogeneous Poisson process,
it is straightforward to simulate sampling times.
We use a \textit{time-transformation} method \citep[pages 98--99]{Cinlar1975},
which, informally, treats the waiting times between events as transformations
of exponential waiting times based on the intensity function following the previous event.
Because the coalescent likelihood is sufficiently similar to an inhomogeneous Poisson process,
we can use a similar time-transformation technique to generate
the coalescent events of simulated genealogies \citep{slatkin1991pairwise}.
We implement these methods for simulating sampling times and coalescent times
in R package \texttt{phylodyn} \citep{Karcher2017}.

In Figure \ref{fig:SimpleCovariatesComparison},
we illustrate BNPR, BNPR-PS, and BNPR-PS with simple covariates
applied to a single simulated genealogy with sampling events distributed
according to log-intensity $1.56 + \gamma(t) - 0.05t$, resulting in 1013 tips, where
$\gamma(t) = \log [N_{e,2,6}(t)]$ and
$N_{e,a,o}(t)$ is a family of functions that approximate seasonal changes
in effective population size, defined as follows:
\begin{equation}
\label{eq:ne}
N_{e,a,o}(t) = \begin{cases} 2 + 18/(1+\exp\{a[3-(t+o \mbox{ (mod 12)})]\}), & \mbox{if } t+o \mbox{ (mod 12) } \leq 6, \\
2 + 18/(1+\exp\{a[3 + (t+o \mbox{ (mod 12)}) - 12]\}) , & \mbox{if } t+o \mbox{ (mod 12) } > 6. \end{cases}
\end{equation}

We see that BNPR (the sampling conditional model) suffers from the kind
of model misspecification induced bias illustrated in \citep{karcher2015quantifying}.
BNPR-PS with no additional covariates beyond $\gamma(t) = \log [N_e(t)]$,
in contrast, suffers even more strongly from a misspecified sampling model.
Table \ref{tab:simple_sim_coefs} shows that the model fails to correctly infer the coefficient of $\gamma(t)$.
This illustrates the care one must take in choosing parameterizations of
the sampling model.
BNPR-PS with simple covariates, $\gamma(t)$ and $-t$, the correctly-specified model,
produces a reconstruction of the effective population trajectory
that is very close to the true trajectory used to simulate the data.
Table \ref{tab:simple_sim_coefs} shows that the true values of the sampling model coefficients
are within 95\% Bayesian credible intervals produced by our inference method
with the correctly specified model.

\begin{table}
\centering
\begin{tabular}{lcrrrr}
\hline
\rowcolor{Gray}
\textbf{Model} & \textbf{Coef} & \textbf{Q0.025} & \textbf{Median} & \textbf{Q0.975} & \textbf{Truth}\\
\hline
$\{\gamma(t)\}$ & $\gamma(t)$ & 1.67 & 1.99 & 2.34 & 1.0\\
\rowcolor{Gray}
$\{\gamma(t), -t\}$ & $\gamma(t)$ & 0.86 & 1.01 & 1.16 & 1.0\\
\rowcolor{Gray}
                               &      $-t$     & 0.040 & 0.047 & 0.053 & 0.050\\
\hline
\end{tabular}
\caption{{\bf Summary of simulated fixed-tree data inference.}
Posterior distribution quantile summaries for
BNPR-PS with no covariates (model: $\{\gamma(t)\}$) and
BNPR-PS with an ordinary covariate (model: $\{\gamma(t), -t\}$).
\label{tab:simple_sim_coefs}}
\end{table}

%

\subsubsection{Direct Inference from Sequence Data}

We simulate several genealogies and DNA sequences from different sampling scenarios
in order to evaluate how well our population reconstruction and parameter inference performs.
Given a sampling model and, optionally, an effective population size trajectory,
we generate sampling times within a \textit{sampling window}.
We generate sampling and coalescent times for a genealogy using
the same time-transformation methods as for our fixed-tree simulations.
We simulate the topology of the genealogy by proceeding backward in time,
adding an active lineage at each sampling time and joining a pair of active lineages
uniformly at random at each coalescent event.
We provide an implementation of this tree-topology simulation method in \texttt{phylodyn}.
We generate simulated sequence alignments using the software SeqGen \citep{rambaut1997seq},
using the Jukes-Cantor 1969 (JC69) \citep{jukes1969evolution} substitution model.
We set the substitution rate to produce an expected 0.9 mutations per site,
in order to produce a sequence alignment with many sites having one mutation,
and some sites having zero or multiple mutations.
For all of our simulations, we use the same seasonal effective population trajectory, $N_{e,2,6}(t)$,
as for our fixed-tree simulations.

\begin{figure}
	\centering
	\includegraphics[width=0.9\textwidth]{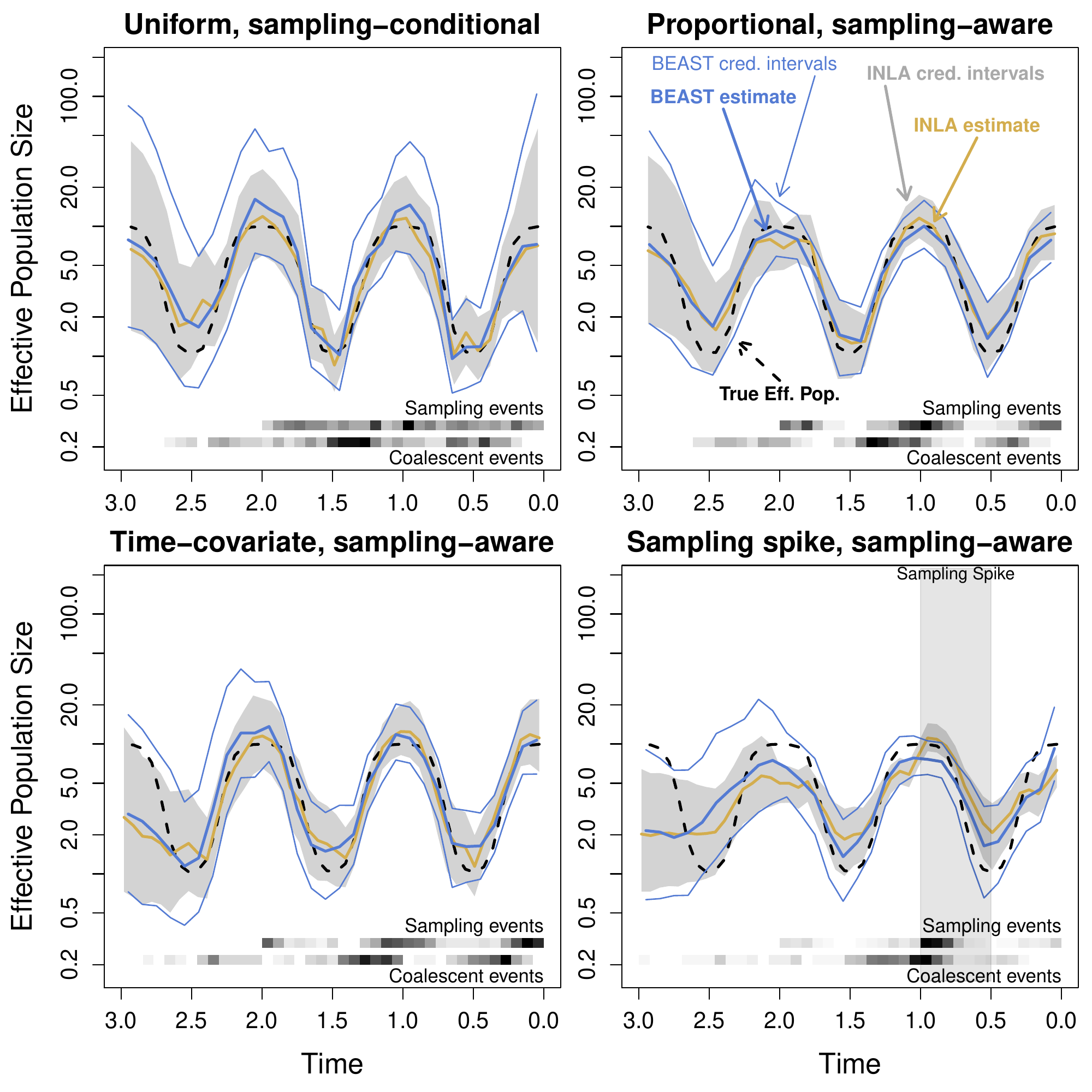}
	\caption{{\bf Effective population size reconstructions for four sequence data simulations,
	all based on the same seasonal effective population size trajectory.}
	\textit{Upper left}: Uniform sampling times, sampling-conditional posterior.
	\textit{Upper right}: Sampling frequency proportional to effective population size,
	sampling-aware posterior.
	\textit{Lower left}: Sampling frequency proportional to effective population times a time-covariate ($\exp(t)$),
	sampling- and covariate-aware posterior.
	\textit{Lower right}: Sampling frequency proportional to effective population size with a sampling spike,
	sampling- and covariate-aware posterior.}
	\label{fig:sequence_simulations}
\end{figure}

First, we simulate a genealogy with 200 tips and
sequence data with 1500 sites and uniform sampling times
and apply both of our sampling-conditional methods.
We apply the INLA-based fixed-tree BNPR from \citep{karcher2015quantifying} to the true genealogy,
and we apply the MCMC-based tree-sampling ESS/BEAST (specified above) to the sequence data.
In Figure \ref{fig:sequence_simulations} (upper left), we compare the truth with
the resulting pointwise posterior medians and credible intervals.
The two methods' results are mutually consistent,
with additional uncertainty in the tree-sampling method (visible in the wider credible intervals)
due to having to estimate the genealogy jointly with other model parameters.
We see similar results comparing BNPR-PS
with SampESS/BEAST in Figure \ref{fig:sequence_simulations} (upper right),
where we sample sequences (1500 sites) with sampling times
generated from an inhomogeneous Poisson process with intensity proportional to effective population size
(log-intensity $2.9 + \gamma(t)$) resulting in 170 samples
and infer using a sampling model with log-intensity $\beta_0 + \beta_1 \gamma(t)$.
We also see similar results in Figure \ref{fig:sequence_simulations} (lower left),
where we add time as an additional covariate and
sample sequences (1500 sites) with log-intensity $3.35 + \gamma(t) - 0.5t$, resulting in 199 samples,
and perform inference using a sampling model with log-intensity $\beta_0 + \beta_1 \gamma(t) + \beta_2 \cdot (-t)$.
Table \ref{tab:sim_coefs} shows that SampESS does a reasonable job at reconstructing the true
model coefficients, though the credible interval for $-t$ includes 0.

We also simulate a genealogy and sequence data (1500 sites) with log-intensity
$1.89 + \gamma(t) + \gamma(t) \cdot 1_{t \in [0.5,1]}$, resulting in 210 samples.
This produces an interval we refer to as a \textit{sampling spike}
which requires the use of an interaction covariate.
Because of design limitations of the R implementation of INLA,
we are limited in how we may implement interaction covariates in BNPR-PS.
Therefore, in Figure \ref{fig:sequence_simulations} (lower right) we plot SampESS/BEAST with
the correct interaction covariate (and a corresponding ordinary covariate)
against BNPR-PS with no covariates.
We see SampESS (with covariates) perform better than BNPR-PS (without covariates)
at reconstructing the correct trajectory.
We also see that our method, using the full covariate model,
with log-intensity $\beta_0 + \beta_1 \gamma(t) + \beta_2 \cdot 1_{t \in [0.5,1]} + \delta_2 \cdot \gamma(t) \cdot 1_{t \in [0.5,1]}$,
produces a 95\% Bayesian credible interval for the coefficient of the ordinary covariate
that contains the true value ($\beta_2 = 0$),
while the true value of the interaction covariate coefficient ($\delta_2 = 1$) is correctly inside
the 95\% Bayesian credible interval produced by SampESS/BEAST.

\begin{table}[ht]
\centering
\begin{tabular}{llrrrr}
\hline
\rowcolor{Gray}
\textbf{Model} & \textbf{Coef} & \textbf{Q0.025} & \textbf{Median} & \textbf{Q0.975} & \textbf{Truth}\\
\hline
$\{\gamma(t)\}$ & $\gamma(t)$ & 0.98 & 1.42 & 2.16 & 1.0\\
\rowcolor{Gray}
$\{\gamma(t), -t\}$ & $\gamma(t)$ & 0.75 & 1.06 & 1.55 & 1.0\\
\rowcolor{Gray}
                               &      $-t$     & -0.06 & 0.44 & 0.94 & 0.5\\
$\{\gamma(t), 1_{t \in [0.5,1]}, 1_{t \in [0.5,1]}\cdot\gamma(t)\}$ & $\gamma(t)$ & 0.72 & 1.26 & 2.14 & 1.0\\
 &   $1_{t \in [0.5,1]}$   & -9.01 & -1.50 & 1.64 & 0.0\\
               & $1_{t \in [0.5,1]}\cdot\gamma(t)$ & 0.13 & 1.75 & 5.75 & 1.0\\
\hline
\end{tabular}
\caption{{\bf Summary of simulated sequence data inference}
Posterior distribution quantile summaries for
SampESS with no covariates (model: $\{\gamma(t)\}$),
SampESS with an ordinary covariate (model: $\{\gamma(t), -t\}$),
and SampESS with both an ordinary and interaction covariate (model: $\{\gamma(t), 1_{t \in [0.5,1]}, 1_{t \in [0.5,1]} \cdot \gamma(t)\}$).\label{tab:sim_coefs}}
\end{table}

\subsection{Seasonal Influenza Example}

\begin{figure}
	\centering
	\includegraphics[width=1.0\textwidth]{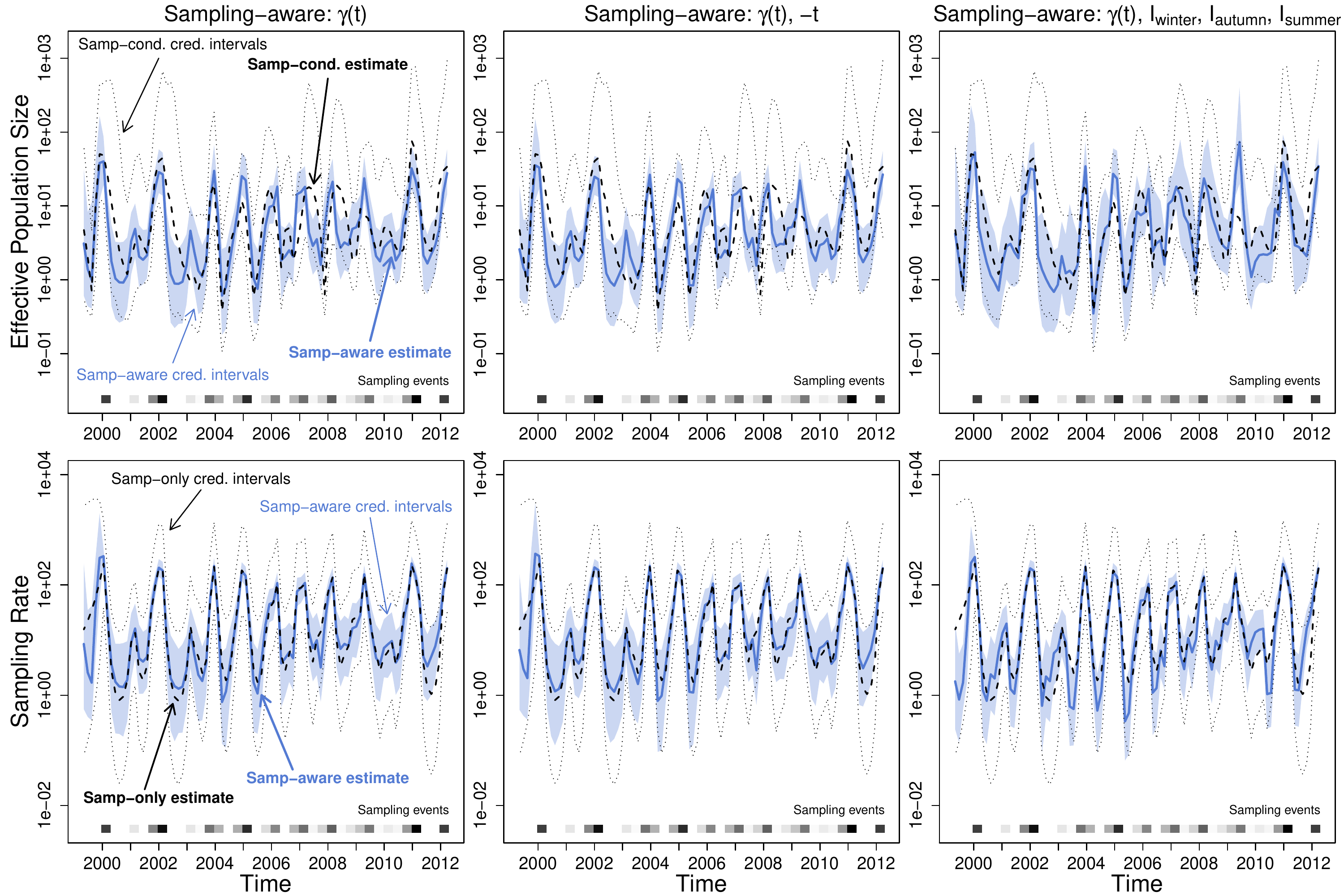}
	\caption{{\bf Effective population size and sampling rate reconstructions for the USA and Canada influenza dataset.}
	\textit{Upper row}: Dashed lines and dotted black lines are
	the pointwise posterior effective population size estimates and
	credible intervals of the sampling-conditional model.
	The blue lines and the light blue regions are
	the pointwise posterior effective population size estimates and
	credible intervals of that column's sampling-aware model.
	\textit{Lower row}: Dashed lines and dotted black lines are
	the pointwise posterior sampling rate estimates and
	credible intervals of a nonparametric sampling-time-only model.
	The blue lines and the light blue regions are
	the pointwise posterior sampling rate estimates and
	credible intervals of that column's sampling-aware model.}
	\label{fig:USACanada}
\end{figure}

We reanalyze the H3N2 regional influenza data for the USA/Canada region
as analyzed with fixed-tree methods in \citep{karcher2015quantifying}.
The data contain 520 sequences aligned to form a multiple sequence alignment
with 1698 sites of the hemagglutinin gene.
This dataset is a subset of the dataset of influenza sequences from around the world analyzed in \citep{zinder2014seasonality}.
We use ESS/BEAST with our tree-sampling MCMC targeting posterior
\eqref{eq:noprefsampling_posterior} to analyze these data
and mark the pointwise posterior median and 95\% credible region in black,
summarized in Figure \ref{fig:USACanada} (upper row).
We observe a seasonal pattern consistent with flu seasons observed in
the temperate northern hemisphere \citep{zinder2014seasonality}.
Our results are also consistent with previous fixed-tree method results
but with larger credible interval widths due to
correctly accounting for genealogical uncertainty in our analysis.

We apply our sampling-aware model SampESS/BEAST to the USA/Canada influenza data,
following the posterior from Equation \eqref{eq:mainposteriorsamp}.
We used several different log-sampling-intensity models.
The simplest one has log-intensity $\beta_0 + \beta_1 \gamma(t)$ (abbreviated $\{\gamma(t)\}$)
and is summarized in Figure \ref{fig:USACanada} (upper left).
We include a time term in one model,
with log-intensity ${\beta_0 + \beta_1 \gamma(t) + \beta_2 \cdot (-t)}$
(abbreviated $\{\gamma(t), -t\}$)
summarized in Figure \ref{fig:USACanada} (upper center).
We use seasonal indicator functions in the final model, defined as,
\begin{align*}
I_{\text{winter}}(t) &= I_{(t \bmod 1.0) \in [0,0.25)},\\
I_{\text{autumn}}(t) &= I_{(t \bmod 1.0) \in [0.25,0.5)},\\
I_{\text{summer}}(t) &= I_{(t \bmod 1.0) \in [0.5,075)},
\end{align*}
with $t$ measured in decimal calendar years (going forward in time).
This results in the log-intensity
$\beta_0 + \beta_1 \gamma(t) +  \beta_2 I_{\text{winter}}(t) +  \beta_3 I_{\text{autumn}}(t) +  \beta_4 I_{\text{summer}}(t)$
(abbreviated $\{\gamma(t), I_{\text{winter}},\allowbreak I_{\text{autumn}},\allowbreak I_{\text{summer}}\}$),
summarized in Figure \ref{fig:USACanada} (upper right).

We summarize the sampling model coefficient results for each model in Table \ref{tab:USACanada_coefs}.
The $\{\gamma(t)\}$ model corresponds to the preferential sampling model of \citet{karcher2015quantifying},
but has noticeably different estimates.
We attribute this to the differences between the fixed-tree
(with a tree inferred using a constant effective population size BEAST model),
INLA-based approach of \citet{karcher2015quantifying},
and the tree-sampling MCMC-based approach of this paper.
We also note that the $\{\gamma(t), -t\}$ model does not perform better
(or even noticeably differently) than the $\{\gamma(t)\}$ model.
The coefficient summary for $\{\gamma(t), -t\}$ bears this out,
because the 95\% Bayesian credible interval for the coefficient for $-t$ contains 0.
This is expected as each year has approximately the same number of sequences, so there should be no exponential growth of sampling intensity. 
We do observe differences in the
$\{\gamma(t), I_{\text{winter}}$, $I_{\text{autumn}}$, $I_{\text{summer}}\}$ model.
The coefficient of $\gamma(t)$ is close to 1.0,
which is the easiest value to interpret under preferential sampling,
suggesting a baseline sampling rate proportional to effective population size.
The coefficients for the indicators suggest increased sampling in the flu season intervals,
as compared to the summer intervals and especially the spring intervals---with
spring treated as a baseline rate without an indicator for the sake of identifiability.
We also fit a model with seasonal indicator covariates and their interactions with the log-effective population size, but do not find any support for including the interaction covariates into the preferential sampling model (see Appendix A.1).

\begin{table}
\centering
\begin{tabular}{llrrr}
\hline
\rowcolor{Gray}
\textbf{Model} & \textbf{Coef} & \textbf{Q0.025} & \textbf{Median} & \textbf{Q0.975}\\
\hline
										$  \{\gamma(t)\}  $ & $\gamma(t)$ & 1.11 & 1.45 & 2.01\\
\rowcolor{Gray}
										$\{\gamma(t), -t\}$ & $\gamma(t)$ & 1.21 & 1.52 & 2.00\\
\rowcolor{Gray}
																&         $-t$       & -0.10 & -0.02 & 0.07\\
     $\{\gamma(t), I_{\text{winter}}, I_{\text{autumn}}, I_{\text{summer}} \}$        &		$\gamma(t)$		& 0.72 & 0.92 & 1.21\\
 & $I_{\text{winter}}$ & 1.91 & 2.79 & 3.83\\
                                                                & $I_{\text{autumn}}$ & 1.88 & 2.85 & 3.85\\
                                                                & $I_{\text{summer}}$ & 0.44 & 1.52 & 2.58\\
\hline
\end{tabular}
\caption{{\bf Summary of USA/Canada influenza data inference.}
Posterior distribution quantile summaries for SampESS with no covariates (model: $\{\gamma(t)\}$),
SampESS with an ordinary covariate (model: $\{\gamma(t), -t\}$),
and SampESS with seasonal indicator covariates (model: $\{\gamma(t),  I_{\text{winter}}, I_{\text{autumn}}, I_{\text{summer}}\}$).\label{tab:USACanada_coefs}}
\end{table}

\begin{figure}
	\centering
	\includegraphics[width=0.9\textwidth]{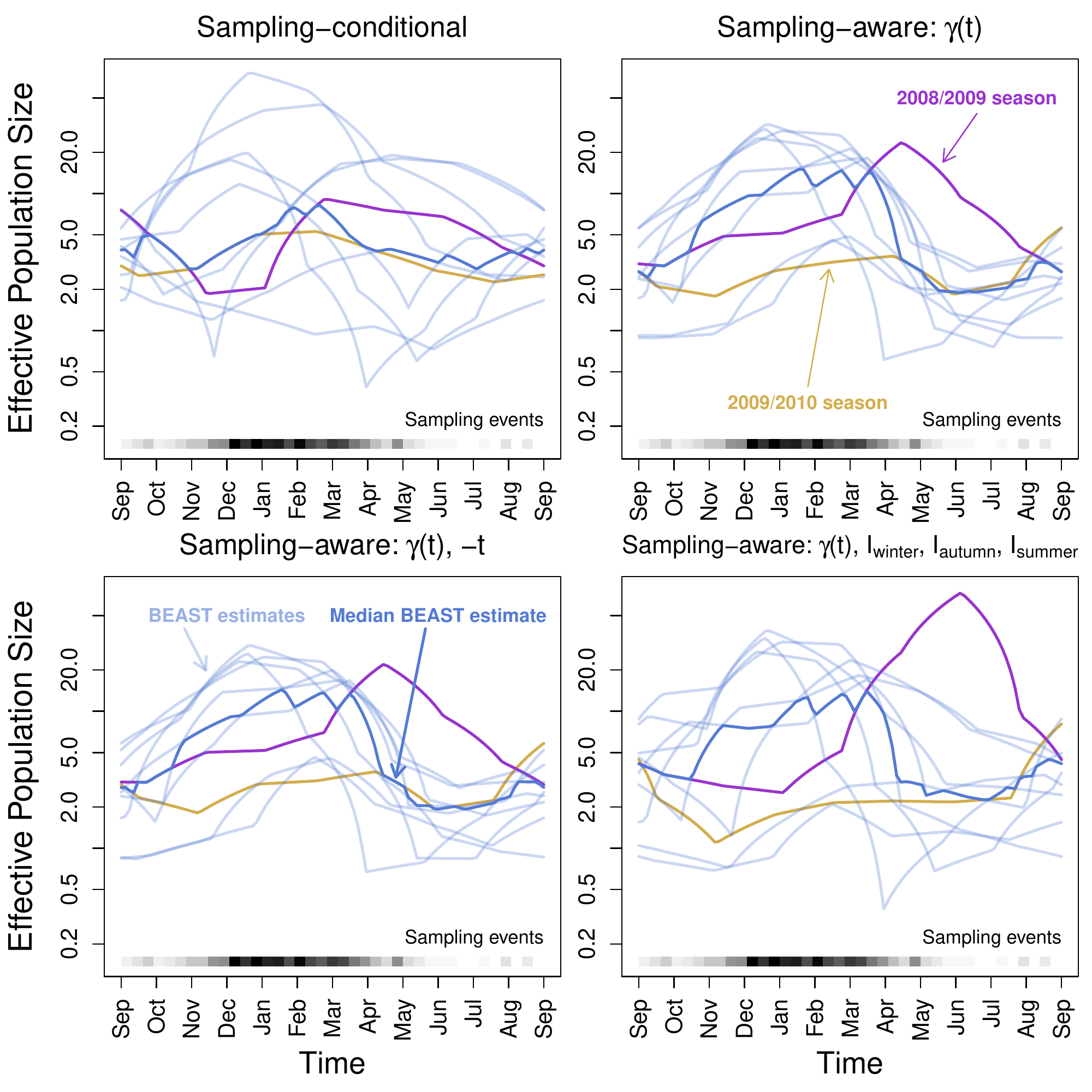}
	\caption{{\bf Effective population size seasonal overlay for the USA and Canada influenza dataset.}
	The light blue lines are the pointwise posterior estimates for each year,
	and the dark blue line is the median annual estimate.
	\textit{Upper left}: Sampling-conditional posterior.
	\textit{Upper right}: Sampling-aware posterior with only log-effective population size $\gamma(t)$ informing the sampling time model.
	\textit{Lower left}: Sampling- and covariate-aware posterior, with $\gamma(t)$ and $-t$.
	\textit{Lower right}: Sampling- and covariate-aware posterior, with $\gamma(t)$ and seasonal indicators $I_{\text{winter}}, I_{\text{autumn}}, I_{\text{summer}}$.}
	\label{fig:trevorograms}
\end{figure}

We observe the seasonality of our estimates of the effective population size trajectory.
In Figure \ref{fig:trevorograms}, we superimpose the twelve years of estimates per model,
and plot the posterior median annual estimate.
We note that the sampling aware models all show increased seasonality compared
to the sampling conditional model.
We also note that the 2008-2009 flu season stands out on the seasonality plot
for having a peak in the summer months of 2009, particularly in the preferential sampling models.
This behavior is most likely due to a misspecification of our model for sampling intensity.
This misspecification is expected given the first documented emergence of
the H1N1 strain in the United States in April of 2009 and the resulting, unaccounted in our model,
increased surveillance of all influenza strains in summer 2009 \citep{CDCH1N1}.
Higher than usual sampling intensity in summer 2009 makes our preferential sampling models conclude
that the effective population size during this time period must be also elevated.
Also, note that the estimated effective population size of H3N2 strain during 2009/2010 flu season
is markedly lower than during most of other seasons.
This is in line with the H1N1 strain successfully competing with the H3N2 strain,
resulting in the lower prevalence of the latter.

To check adequacy of the preferential sampling models, we compare the posterior distributions of sampling intensities obtained via our  BEAST implementation of BNPR-PS with
a nonparametric INLA-based estimate of the sampling rate
(using a method similar to BNPR-PS without the coalescent likelihood or covariates).
Figure \ref{fig:USACanada} (lower row) shows the comparison.
The methods produce very similar estimates,
with the BEAST/MCMC methods having thinner credible intervals
due to incorporating additional information from the coalescent likelihood.
We also performed posterior predictive checking, described in Appendix B.1, but found that this method lacked power to discriminate among preferential sampling models (see Appendix B.2.2).

\subsection{Ebola Outbreak}
\label{ebola_results}
Next, we analyze a subset of the Ebola virus sequences  arising from
the recent Western Africa Ebola outbreak (as collated in  \citep{dudas2017virus}).
The data consist of 1610 aligned whole genomes,
collected from mid-2014 to mid-2015.
The resulting alignment has 18,992 sites.
The dataset represents over 5\% of known cases of Ebola detected during that outbreak,
providing an unprecedented insight into the epidemiological dynamics of an Ebola outbreak.
We consider two subsets of the data, corresponding to the samples from Sierra Leone and Liberia.
For Sierra Leone, we subsampled 200 sequences, chosen uniformly at random out of 1010 samples
for computational tractability.
For Liberia, we use the entire collection of 205 sequences obtained from infected individuals in this country.

\begin{figure}
	\centering
	\includegraphics[width=0.99\textwidth]{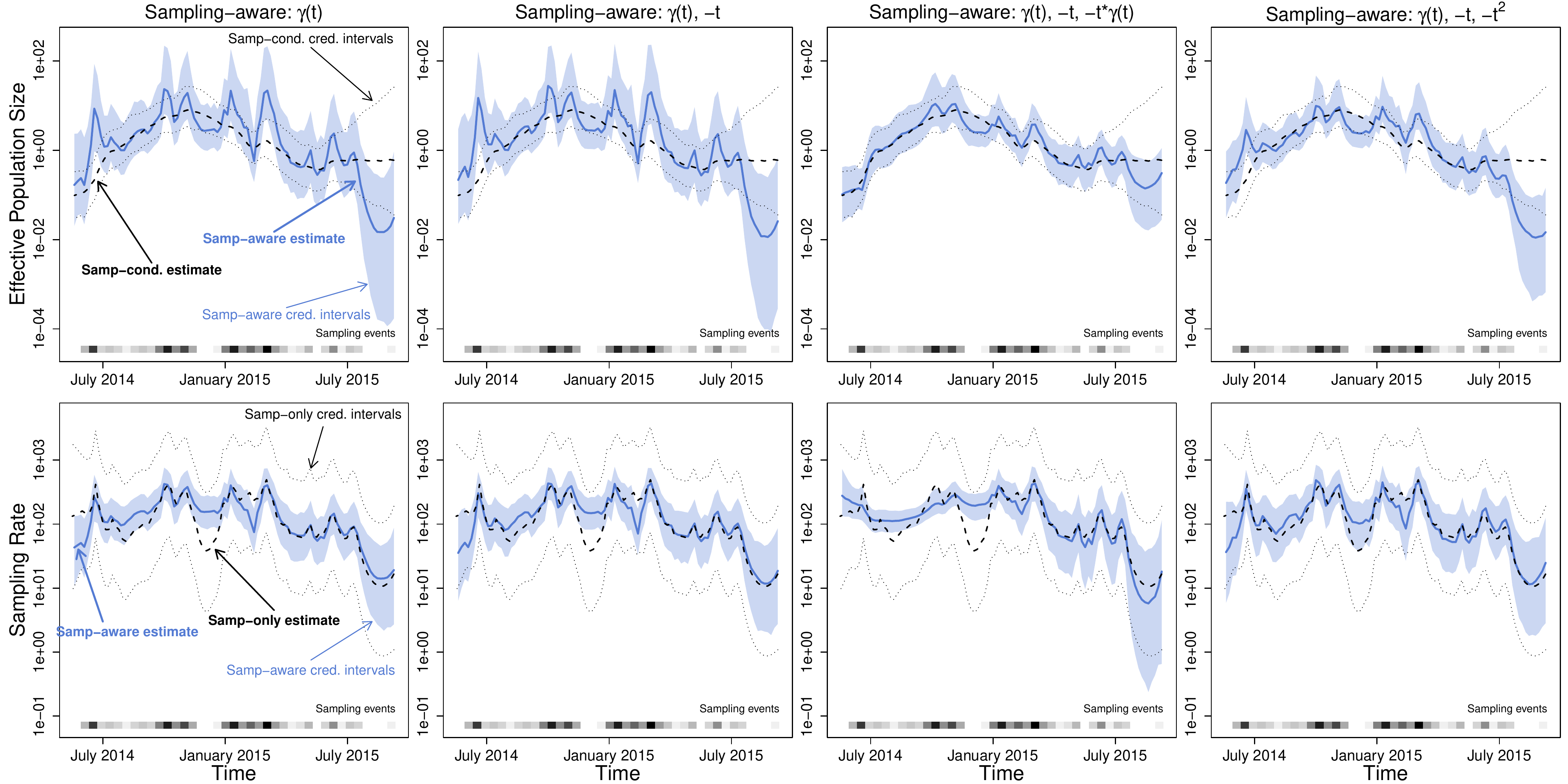}
	\caption{{\bf Effective population size and sampling rate reconstructions for the Sierra Leone Ebola dataset.}
	\textit{Upper row}: Dashed lines and dotted black lines are
	the pointwise posterior effective population size estimates and
	credible intervals of the sampling-conditional model.
	The blue lines and the light blue regions are
	the pointwise posterior effective population size estimates and
	credible intervals of that column's sampling-aware model.
	\textit{Lower row}: Dashed lines and dotted black lines are
	the pointwise posterior sampling rate estimates and
	credible intervals of a nonparametric sampling-time-only model.
	The blue lines and the light blue regions are
	the pointwise posterior sampling rate estimates and
	credible intervals of that column's sampling-aware model.}
	\label{fig:Ebola}
\end{figure}

We begin by applying ESS/BEAST method with no preferential sampling to the Sierra Leone dataset.
We use MCMC to target our tree-sampling posterior from Equation \eqref{eq:noprefsampling_posterior}
and depict the pointwise posterior median effective population curve, $N_e(t)$,
with a black dashed line and its corresponding 95\% credible region boundaries with black doted lines,
shown in all panels of the first row of Figure \ref{fig:Ebola}.
The resulting effective population size trajectory visually
resembles a typical epidemic trajectory of prevalence or incidence that peaks in Autumn of 2014.
Next, we apply our sampling-aware model SampESS/BEAST to the Ebola data,
targeting with MCMC the posterior from Equation \eqref{eq:mainposteriorsamp}.
We use several different log-sampling-intensity models.
The simplest model, abbreviated as $\{\gamma(t)\}$,
has log-sampling-intensity $\beta_0 + \beta_1 \gamma(t)$.
We include a $t$ term in our next sampling model, abbreviated as $\{\gamma(t), -t\}$,
with log-intensity ${\beta_0 + \beta_1 \gamma(t) + \beta_2 \cdot (-t)}$.
This model postulates that even if the effective population size remains constant,
the sampling intensity is growing or declining exponentially.
We make $-t$ an interaction covariate as well in the next model,
abbreviated $\{\gamma(t), -t, -t \cdot \gamma(t)\}$,
resulting in the log-sampling-intensity
$\beta_0 + \beta_1 \gamma(t) +  \beta_2 \cdot (-t) +  \delta_2 \gamma(t) \cdot (-t)$.
For the final model, we include $-t$ and $-t^2$ as ordinary covariates,
abbreviated $\{\gamma(t), -t, -t^2\}$,
with log-sampling-intensity ${\beta_0 + \beta_1 \gamma(t) + \beta_2 \cdot (-t)} + \beta_3 \cdot (-t^2)$.
The resulting posterior distribution summaries of the effective population size trajectory
are shown in blue in the upper row of Figure \ref{fig:Ebola}.

Having concluded that posterior predictive checks are underpowered in our setting (see Appendix B.2.3), we judge preferential sampling model fit using lessons learned during simulation studies conducted by \citet{karcher2015quantifying}. 
One important observation made by  \citet{karcher2015quantifying} is that ignoring preferential sampling produces little bias when estimating $N_e(t)$ when population size is increasing. Even during the $N_e(t)$ declines the bias is relatively small. 
Therefore, if the sampling model is not misspecified, we expect conditional and sampling aware models to differ mostly in the width of their credible intervals, and not in posterior medians of $N_e(t)$.
According to this metric, the $\{\gamma(t), -t, -t \cdot \gamma(t)\}$ and $\{\gamma(t), -t, -t^2\}$ models show reasonable agreement with the conditional conditional model when reconstructing $N_e(t)$ (see first row of Figure~\ref{fig:Ebola}), with the $\{\gamma(t), -t, -t \cdot \gamma(t)\}$ model performing the best. 
Another measure of goodness of fit is agreement of nonparametric estimate of sampling intensity and the one produced by preferential sampling models.
By this metric, the quadratic model $\{\gamma(t), -t, -t^2\}$ outperforms the interaction model $\{\gamma(t), -t, -t \cdot \gamma(t)\}$.
The main difference between the interaction and quadratic models is that the interaction model suggests that the strength of preferential sampling (power of $N_e(t)$) was linearly increasing over time, 
while the quadratic model points to a quadratically time varying multiplicative factor in front of  $N_e(t)$.
Parameter estimates of all models fit to the Sierra Leone data are summarized in Table \ref{tab:Ebola_coefs}.
\par
As another way to compare adequacy of preferential sampling models, we overlay our reconstructed effective population size trajectories and Ebola weekly incidence time series. 
We use a sum of confirmed and probable case counts from the supplementary data of  \citet{dudas2017virus}.
Assuming a susceptible-infectious-removed (SIR) model, an approximate structured coalescent model, and taking into account the fact that the number of susceptible individuals never decreased appreciably during the Ebola outbreak, we can interpret effective population size as a quantity proportional to Ebola incidence \citep{volz2009phylodynamics, frost2010viral}.
Figure~\ref{fig:Ebola_incidence_Sierra_Leone} shows aligned incidence time series and posterior summaries of effective population size for all considered coalescent models. 
All models produce reasonable agreement between incidence and effective population size during the increase of incidence. However, the end of the outbreak is captured better by preferential sampling models, with the quadratic model $\{\gamma(t), -t, -t^2\}$ outperforming the other preferential sampling models.

\begin{table}[ht]
\centering
\begin{tabular}{llrrr}
\hline
\rowcolor{Gray}
\textbf{Model} & \textbf{Coef} & \textbf{Q0.025} & \textbf{Median} & \textbf{Q0.975}\\
\hline
	$\{\gamma(t)\}$ & $\gamma(t)$ & 0.28 & 0.46 & 0.71\\
\rowcolor{Gray}
$\{\gamma(t), -t\}$ & $\gamma(t)$ & 0.30 & 0.49 & 0.83\\
\rowcolor{Gray}
							&		 $-t$		   & -0.58 & 0.27 & 1.46\\
$\{\gamma(t), -t, -t\cdot\gamma(t)\}$  & $\gamma(t)$  & 1.01 & 1.76 & 3.32\\
 &		$-t$		& -0.88 & 0.18 & 1.02\\
					& $-t\cdot\gamma(t)$ & 0.89 & 1.65 & 3.11\\
\rowcolor{Gray}
	$\{\gamma(t), -t, -t^2\}$ & $\gamma(t)$ & 0.47 & 1.00 & 1.80\\
	\rowcolor{Gray}
	& $-t$ & 2.02 & 9.05 & 20.63\\
	\rowcolor{Gray}
	& $-t^2$ & -13.08 & -5.58 & -1.09\\
\hline
\end{tabular}
\caption{{\bf Summary of Sierra Leone Ebola sequence data inference.}
Posterior distribution quantile summaries for SampESS with no covariates (model: $\{\gamma(t)\}$),
SampESS with an ordinary covariate (model: $\{\gamma(t), -t\}$),
SampESS with both an ordinary and interaction covariate (model: $\{\gamma(t), -t, -t \cdot \gamma(t)\}$),
and SampESS with linear and quadratic ordinary covariates (model: $\{\gamma(t), -t, -t^2\}$).\label{tab:Ebola_coefs}}
\end{table}

\begin{table}[ht]
\centering
\begin{tabular}{llrrr}
\hline
\rowcolor{Gray}
\textbf{Model} & \textbf{Coef} & \textbf{Q0.025} & \textbf{Median} & \textbf{Q0.975}\\
\hline
	$\{\gamma(t)\}$ & $\gamma(t)$ & 0.53 & 0.78 & 1.20\\
\rowcolor{Gray}
$\{\gamma(t), -t\}$ & $\gamma(t)$ & 0.53 & 0.81 & 1.23\\
\rowcolor{Gray}
							&		 $-t$		   & -3.39 & -0.74 & 1.84\\
$\{\gamma(t), -t, -t\cdot\gamma(t)\}$  & $\gamma(t)$  & -0.07 & 0.40 & 1.51\\
 &		$-t$		& -3.26 & -0.31 & 2.67\\
					& $-t\cdot\gamma(t)$ & -2.98 & -1.21 & 1.20\\
\hline
\end{tabular}
\caption{{\bf Summary of Liberia Ebola sequence data inference.}
Posterior distribution quantile summaries for SampESS with no covariates (model: $\{\gamma(t)\}$),
SampESS with an ordinary covariate (model: $\{\gamma(t), -t\}$),
and SampESS with both an ordinary and interaction covariate (model: $\{\gamma(t), -t, -t \cdot \gamma(t)\}$).\label{tab:Liberia_coefs}}
\end{table}

\begin{figure}
	\centering
	\includegraphics[width=0.9\textwidth]{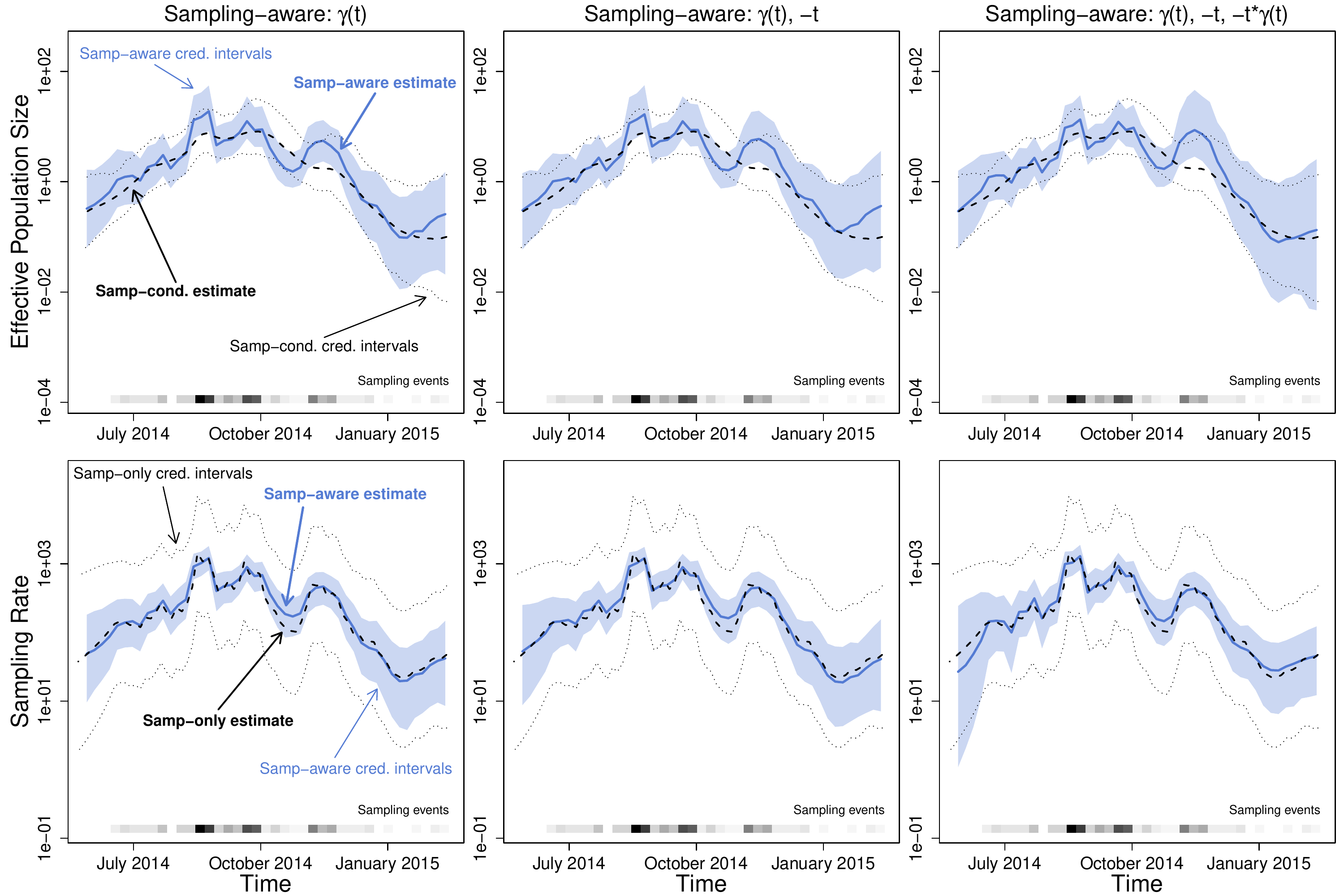}
	\caption{{\bf Effective population size and sampling rate reconstructions for the Liberia Ebola dataset.}
	\textit{Upper row}: Dashed lines and dotted black lines are
	the pointwise posterior effective population size estimates and
	credible intervals of the sampling-conditional model.
	The blue lines and the light blue regions are
	the pointwise posterior effective population size estimates and
	credible intervals of that column's sampling-aware model.
	\textit{Lower row}: Dashed lines and dotted black lines are
	the pointwise posterior sampling rate estimates and
	credible intervals of a nonparametric sampling-time-only model.
	The blue lines and the light blue regions are
	the pointwise posterior sampling rate estimates and
	credible intervals of that column's sampling-aware model.}
	\label{fig:Liberia}
\end{figure}

We apply the same models to the Liberia Ebola dataset,
summarized across the upper row of Figure \ref{fig:Liberia} and in Table \ref{tab:Liberia_coefs}.
We note that the $\{\gamma(t)\}$ and $\{\gamma(t), -t\}$ models perform very similarly,
but the $\{\gamma(t), -t\}$ model has slightly wider pointwise credible intervals in places.
This is consistent with the coefficients, as the credible interval for the $-t$ term contains 0.
The $\{\gamma(t), -t, -t \cdot \gamma(t)\}$ model has even wider pointwise credible intervals,
and the credible intervals for the coefficients all contain 0.
This suggests that in Liberia, of the three sampling-aware models,
the sampling model most consistent with the data is simple preferential sampling.
We also note that in the $\{\gamma(t)\}$ model,
the median estimate for the coefficient for $\gamma(t)$ is close to 1.0,
suggesting direct proportional sampling.
\par
As in the previous section, we compare the sampling rates we derive from our BEAST runs to
a nonparametric INLA-based estimate of the sampling rate.
Figures \ref{fig:Ebola} (lower row) and \ref{fig:Liberia} (lower row) show the comparisons.
The two methods produce very similar estimates,
and again the sampling-aware methods have thinner credible intervals
due to incorporating additional information from the coalescent likelihood.
\par
All coalescent models produce reasonable agreement between estimated effective population size trajectories and Ebola incidence time series in Libera (see Appendix Fig~\ref{fig:Ebola_incidence_Liberia}).
The model without preferential sampling looks the best in this comparison, mostly because the incidence curve does not support multiple ``ups'' and ``downs" in effective population size trajectories estimated under the preferential sampling models.

\section{Discussion}


Currently, few phylodynamic methods incorporate sampling time models
in order to address model misspecification and take advantage of the additional
information contained in sampling times in preferential sampling contexts.
Even fewer methods implement sampling time models by appropriately integrating
over genealogies relating the sampled genetic sequences and performing inference directly
from these sequence data.
We extend previous sampling time models to incorporate
time-varying covariates in order to allow the sampling model
to be more flexible under different scientific circumstances.
We implement this sampling time model into the MCMC software BEAST,
and also implement an elliptical slice sampler into BEAST for efficient MCMC draws of grid-based
effective population size parameterizations.


However, the additional flexibility of the sampling time model comes with additional uncertainty around which set of covariates is the best one for a given scientific context.
Unfortunately, posterior predictive checks \citep{gelman1996posterior} --- taking
a model with a posterior sample of parameters estimated from data,
using the same model and estimated parameters to simulate new data,
and comparing the observed data and the simulated data --- lacked sufficient power to discriminate between models in all of our applications.
Including a residual term to absorb sampling times model misspecification can be a fruitful avenue of future research. 
However, care will need to be taken to preserve parameter identifiability. 


Another approach to extending and increasing the flexibility of the sampling model
is to decouple the fixed temporal relationship between effective population size
and sampling intensity.
Introducing an estimated lag parameter to the sampling time model
would allow for cause-and-effect phenomena and delays to be accounted for
within the model.
Incorporating an estimated lag parameter would also allow for an additional avenue
of model verification.
Under most imaginable circumstances,
if there is a relationship between the effective population size and sampling frequency,
changes to the population size would effect sampling frequency with zero or positive delay.
Estimating a credibly negative lag would be a possible indicator that some
element of the model or data is worth re-examining.


In terms of flexibility, the ideal sampling time model would be
a separate Gaussian latent field distinct from the (log) effective population size.
However, methods for primarily phylodynamic inference with this feature
would suffer from severe identifiability problems.
One approach that would retain most of the flexibility of the separate Gaussian field
while also retaining the identifiability of the original model
would be to model the (log) effective population size and sampling intensities
as \textit{correlated} Gaussian processes.
Estimating the correlation parameter between the two processes would
allow for estimation of the preferential sampling strength.


\section*{Acknowledgments}
We are grateful to Trevor Bedford for hist suggestions that greatly improved our seasonal Influenza data analysis.
M.K., M.A.S, and V.N.M.~were supported by the NIH grant R01 AI107034.
M.A.S.~was supported by the NSF grant DMS 1264153 and NIH grant R01 LM012080.
V.N.M.\ was supported by the NIH grant U54 GM111274.

\FloatBarrier

\bibliography{../bib/phylo}

\clearpage

\appendix
\raggedbottom\sloppy

\pagenumbering{arabic}
\renewcommand*{\thepage}{A-\arabic{page}}

\renewcommand{\thefigure}{A-\arabic{figure}}
\setcounter{figure}{0}

\renewcommand{\thetable}{A-\arabic{table}}
\setcounter{table}{0}

\section{Appendix: Additional Sequence Data Results}
\label{sec:AppendixA}

\subsection{Seasonal Influenza}

\begin{figure}
	\centering
	\includegraphics[width=0.4\textwidth]{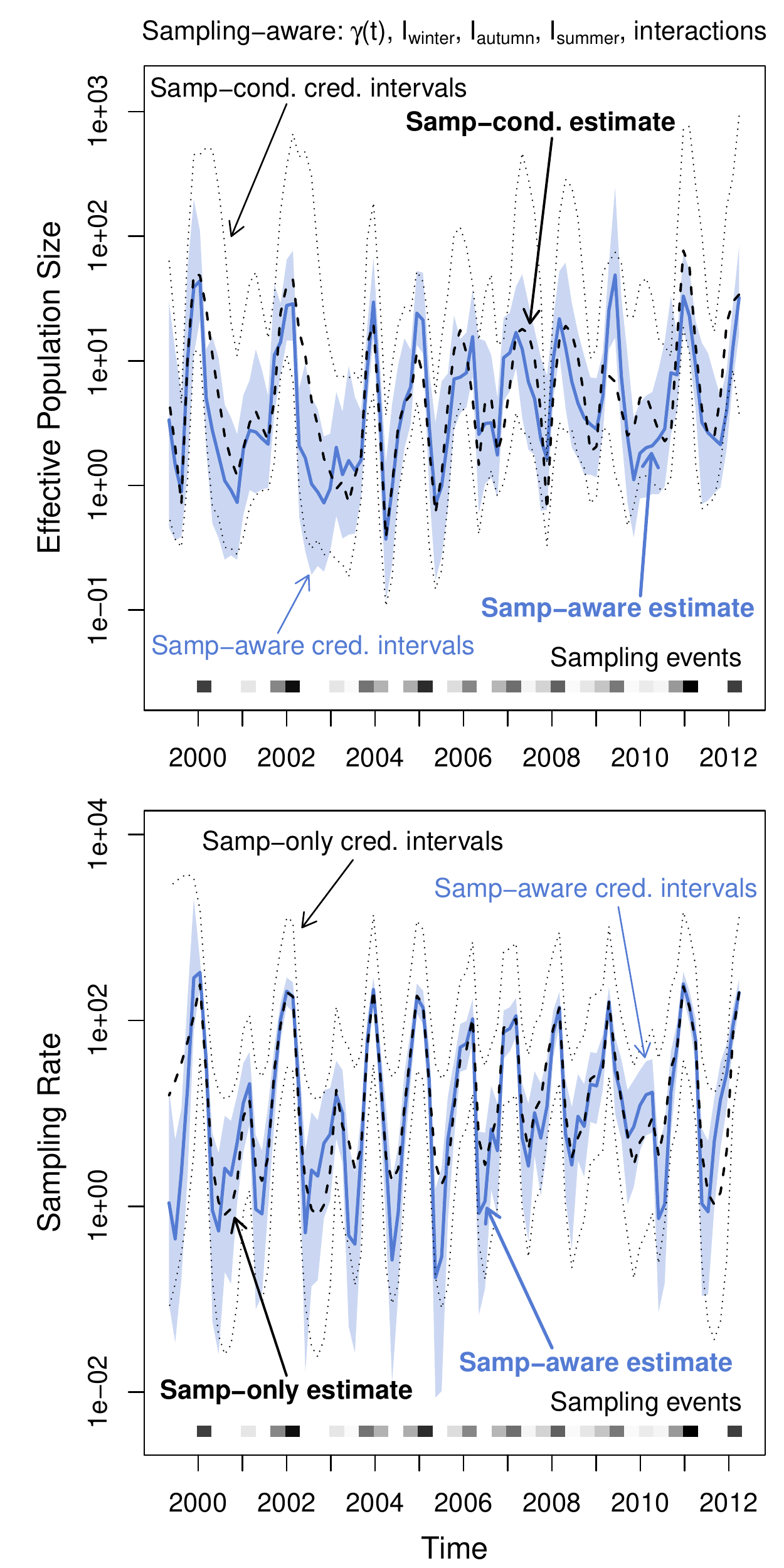}
	\caption{{\bf Effective population size and sampling rate reconstructions for the USA and Canada influenza dataset.}
	\textit{Upper row}: Dashed lines and dotted black lines are 
	the pointwise posterior effective population size estimates and 
	credible intervals of the sampling-conditional model.
	The blue line and the light blue region are 
	the pointwise posterior effective population size estimates and 
	credible intervals of that column's sampling-aware model.
	\textit{Lower row}: Dashed lines and dotted black lines are 
	the pointwise posterior sampling rate estimates and 
	credible intervals of a nonparametric sampling-time-only model.
	The blue line and the light blue region are 
	the pointwise posterior sampling rate estimates and 
	credible intervals of that column's sampling-aware model.}
	\label{fig:USACanada_appendix}
\end{figure}

We consider one additional model for the USA/Canada influenza data with log-intensity,
\begin{align*}
\beta_0 + \beta_1 \gamma(t) &+  \beta_2 I_{\text{winter}}(t) +  \beta_3 I_{\text{autumn}}(t) +  \beta_4 I_{\text{summer}}(t)\\
& + \delta_2 I_{\text{winter}}(t) \cdot \gamma(t) +  \delta_3 I_{\text{autumn}}(t) \cdot \gamma(t) +  \delta_4 I_{\text{summer}}(t) \cdot \gamma(t),
\end{align*}
abbreviated $ \{ \gamma(t)$, $I_{\text{winter}}$, $I_{\text{autumn}}$, $I_{\text{summer}}$, 
$I_{\text{winter}} \cdot \gamma(t)$, $I_{\text{autumn}} \cdot \gamma(t)$, $I_{\text{summer}} \cdot \gamma(t) \} $,
or more succinctly as $ \{ \gamma(t)$, $I_{\text{winter}}$, $I_{\text{autumn}}$, $I_{\text{summer}}$, $\text{interactions}\}$.
The results are summarized in Figure \ref{fig:USACanada_appendix} and Table \ref{tab:USACanada_coefs_appendix}.
We see that only the coefficients for $\gamma(t)$, $I_{\text{winter}}$, and $I_{\text{autumn}}$
have credible intervals that do not contain zero,
suggesting that additional terms are not necessary.

\begin{table}[b]
\centering
\begin{tabular}{llrrr}
\hline
\rowcolor{Gray}
\textbf{Model} & \textbf{Coef} & \textbf{Q0.025} & \textbf{Median} & \textbf{Q0.975}\\
\hline
     $\{\gamma(t), I_{\text{winter}}, I_{\text{autumn}}, I_{\text{summer}},\text{interactions}\}$        &		$\gamma(t)$		& 0.64 & 1.20 & 1.97\\
 & $I_{\text{winter}}$ & 1.83 & 3.48 & 5.58\\ 
& $I_{\text{autumn}}$ & 1.31 & 3.16 & 5.28\\
 & $I_{\text{summer}}$ & -0.15 & 2.08 & 4.52\\
  & $I_{\text{winter}} \cdot \gamma(t)$ & -1.08 & -0.29 & 0.34\\ 
& $I_{\text{autumn}} \cdot \gamma(t)$ & -1.00 & -0.14 & 0.53\\
 & $I_{\text{summer}} \cdot \gamma(t)$ & -1.24 & -0.24 & 0.92\\

\hline
\end{tabular}
\caption{{\bf Summary of USA/Canada influenza data inference.}
Posterior distribution quantile summaries for SampESS with seasonal indicator and interaction covariates 
(model: $\{\gamma(t),  I_{\text{winter}}, I_{\text{autumn}}, I_{\text{summer}}$,$\text{interactions}\}$).}
\label{tab:USACanada_coefs_appendix}
\end{table}

\subsection{Ebola Outbreak}

\begin{figure}
	\centering
	\includegraphics[width=0.8\textwidth]{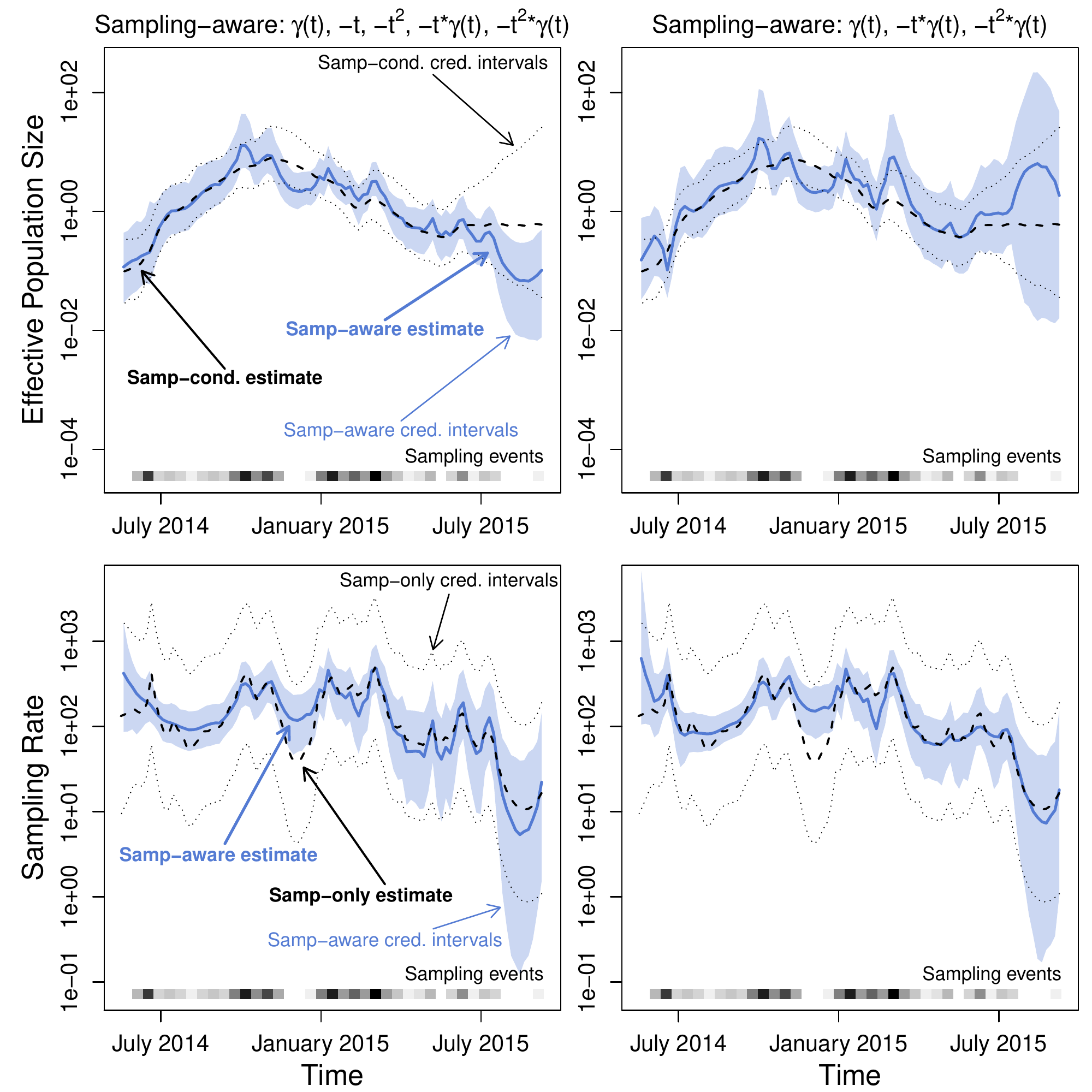}
	\caption{{\bf Effective population size and sampling rate reconstructions for the Sierra Leone Ebola dataset.}
	\textit{Upper row}: Dashed lines and dotted black lines are 
	the pointwise posterior effective population size estimates and 
	credible intervals of the sampling-conditional model.
	The blue line and the light blue region are 
	the pointwise posterior effective population size estimates and 
	credible intervals of that column's sampling-aware model.
	\textit{Lower row}: Dashed lines and dotted black lines are 
	the pointwise posterior sampling rate estimates and 
	credible intervals of a nonparametric sampling-time-only model.
	The blue line and the light blue region are 
	the pointwise posterior sampling rate estimates and 
	credible intervals of that column's sampling-aware model.}
	\label{fig:Ebola_appendix}
\end{figure}

We consider three additional models for our subsample of 200 sequences from
the Sierra Leone Ebola outbreak data with log-intensities,
\begin{align*}
\beta_0 + \beta_1 \gamma(t) &+  \beta_2 \cdot (-t) +  \beta_3 \cdot (-t^2) \\
&+ \delta_2 \gamma(t) \cdot (-t) +  \delta_3 \gamma(t) \cdot (-t^2), \; \text{and} \\
\beta_0 + \beta_1 \gamma(t) &+ \delta_2 \gamma(t) \cdot (-t) +  \delta_3 \gamma(t) \cdot (-t^2),
\end{align*}
abbreviated as 
$\{\gamma(t), -t, -t^2$, $-t \cdot \gamma(t)$, $-t^2  \cdot \gamma(t)\}$,
and $\{\gamma(t)$, $-t \cdot \gamma(t)$, $-t^2\cdot\gamma(t)\}$, respectively.
The results are summarized in Figure \ref{fig:Ebola_appendix} and Table \ref{tab:Ebola_coefs_appendix}.
We see that the coefficients for $\gamma(t)$, $-t$, and $-t^2$ tend to
have credible intervals that do not contain zero 
(except for the interaction-only model $\{\gamma(t)$, $-t \cdot \gamma(t)$, $-t^2\cdot\gamma(t)\}$),
but the other terms do not, suggesting that the additional terms are not necessary.

\begin{table}[ht]
\centering
\begin{tabular}{llrrr}
\hline
\rowcolor{Gray}
\textbf{Model} & \textbf{Coef} & \textbf{Q0.025} & \textbf{Median} & \textbf{Q0.975}\\
\hline
$\{\gamma(t), -t, -t^2, -t  \cdot \gamma(t), -t^2  \cdot \gamma(t)\}$ & $\gamma(t)$ & 0.71 & 2.20 & 4.69\\
							&		 $-t$		   & 1.21 & 9.75 & 20.29\\
							&		 $-t^2$		   & -12.67 & -6.00 & -0.79\\
							&		 $-t \cdot \gamma(t)$		   & -2.72 & 0.95 & 6.49\\
							&		 $-t^2 \cdot \gamma(t)$		   & -2.64 & 0.72 & 3.26\\
\rowcolor{Gray}
$\{\gamma(t), -t \cdot \gamma(t), -t^2\cdot\gamma(t)\}$  & $\gamma(t)$  & -3.00 & -1.39 & 2.08\\
\rowcolor{Gray}
 &		$-t \cdot \gamma(t)$		& -11.30 & -6.59 & 2.00\\ 
 \rowcolor{Gray}
					& $-t^2\cdot\gamma(t)$ & -0.16 & 4.90 & 8.16\\
\hline
\end{tabular}
\caption{{\bf Summary of Sierra Leone Ebola sequence data inference.}
Posterior distribution quantile summaries for SampESS with models:
$\{\gamma(t), -t, -t^2$, $-t \cdot \gamma(t), -t^2  \cdot \gamma(t)\}$,
and $\{\gamma(t), -t \cdot \gamma(t), -t^2\cdot\gamma(t)\}$.}
\label{tab:Ebola_coefs_appendix}
\end{table}

\clearpage

\pagenumbering{arabic}
\renewcommand*{\thepage}{B-\arabic{page}}

\renewcommand{\thefigure}{B-\arabic{figure}}
\setcounter{figure}{0}

\renewcommand{\thetable}{B-\arabic{table}}
\setcounter{table}{0}

\section{Appendix: Model Checks and Model Selection}
\label{sec:AppendixB}
 

\subsection{Methods}

\subsubsection{Transformed Exponentials}

Suppose random variable $X \sim \Exp(1)$,
and thus its PDF is $f_X(x) = \exp(-x)$.
Define ${g_\lambda(u) = \int_{0}^u {\lambda(t)dt}}$ for
nonnegative $\lambda(\cdot)$ integrable on $[0,\infty)$.
Then $g_\lambda(u)$ is monotonic nondecreasing,
so $g_\lambda^{-1}(\cdot)$ is well-defined almost everywhere.
If we let $U = g_\lambda^{-1}(X)$,
then the PDF of $U$ is $f_U(u) = \lambda(u) \exp(-\int_{0}^u {\lambda(t)dt})$.

We then have two useful results.
If we wish to sample $U$,
we may do so by sampling an
$\Exp(1)$ random variable $X$,
then apply the transformations $U = g_{\lambda}^{-1}(X)$,
which will result in the desired distribution.
There generally is not an explicit, closed-form solution for $g^{-1}(\cdot)$,
but it can be implicitly solved using root-finding methods and,
if necessary, numerical integration.
Conversely, if we wish to recover the original $\Exp(1)$ random variable $X$
from $U$,
we can apply the transformations $X = g_{\lambda}(U)$.

\subsubsection{Heterochronous Coalescent Time Transformation}


Consider the heterochronous coalescent model, as presented in 
Section \ref{sec:methods} of the main text. 
\citet{griffiths1994ancestral} show that for isochronous data,
the sequence of coalescent events of a genealogy (and allowing variable effective population size) 
is a continuous time Markov chain and that the function $A_n(t)$, 
representing the number of distinct ancestors at time $t$ and called the \textit{ancestral process},
is a pure death process starting at value $n$ at time $0$ and decreasing by one at every coalescent event
proceeding into the past.

We seek to extend this framework to allow heterochronous genealogies as well.
Consider a Wright-Fisher population with population $N(i)$, $i$ generations in the past.
We assume that sampled individuals cannot be ancestors to future sampled individuals,
so if we sample an individual at generation $i$,
we segregate that individual from the other $N(i)$ individuals in the population
until the sampled individual ``selects'' an ancestor in generation $i+1$,
at which point the usual Wright-Fisher process proceeds until another individual is sampled farther in the past.
Suppose we have a fixed schedule of $n$ individuals sampled at generations $g_1 \leq g_2 \leq \ldots \leq g_n$,
and we consider any particular generation $i$,
having counted $k$ coalescent events between generation $0$ and generation $i$.
Let $b_i = \sum_{i=1}^n 1_{[g_i > i]}$ represent the number of individuals that are sampled 
farther into the past than generation $i$.
In an isochronous scenario, $b_i$ would be $0$ for all $i$,
and the number of distinct lineages at generation $i$ would be $n-k$.
However, here we suppose that $b_i > 0$.
We see that if there are no individuals sampled at generations $i$ or $i+1$,
then this iteration of the Wright-Fisher process is identical to an iteration of 
an isochronous Wright-Fisher process with the same population and $n-k-b_i$ distinct lineages.
If there is an individual sampled at generation $i+1$, the outcome is the same
since we can safely ignore the (segregated) sampled individual until iterating from generation $i+1$ to $i+2$.
If there is an individual sampled at generation $i$, 
then we consider the (segregated) sampled individual to be an additional distinct lineage,
but we see the iteration still behaves as if it were an iteration of an isochronous Wright-Fisher process 
with $n-k-b_i$ distinct lineages.

We now switch to continuous time, applying our heterochronous distinct lineage counts 
into the results from \citep{griffiths1994ancestral}.
Let $b(t) = \sum_{i=1}^n 1_{[s_i > t]}$ be the count of samples that occur farther into the past than time $t$.
Let $B_n(t) = n - k(t)$, where $k(t)$ is the number of coalescent events between time $0$ and time $t$.
Under isochronous sampling, $B_n(t) = A_n(t)$ is the ancestral process.
Under heterochronous sampling, $B_n(t)$ is merely the pure death process that is directly analogous to $A_n(t)$.
Substituting our results from the heterochronous Wright-Fisher process into the key results
reveals the transition rates for $B_n(t)$,
\[
\Pr(B_n(t+h) = j \mid B_n(t) = i) = \begin{cases}
\binom{i - b(t)}{2}\frac{1}{N_e(t)}h + o(h), & j = i - 1 \\
1 - \binom{i - b(t)}{2}\frac{1}{N_e(t)}h + o(h), & j = i \\
0 & \text{otherwise},
 \end{cases}
\]
and the joint density for the Markov chain of coalescent events,
\[
\Pr(\mathbf{g} \mid N_e(t), \mathbf{s}) = 
         \prod_{k=2}^n {\left[\lambda_k(t_{k-1}) \exp\left(-\int_{t_k}^{t_{k-1}} {\lambda_k(t)dt}\right)\right]},
\]
where $\lambda_k(t) = \binom{k - b(t)}{2}\frac{1}{N_e(t)}$.

Following the results from \citep{griffiths1994ancestral}, 
we note that the terms in the product are in the form of transformed exponentials,
and can be sampled by transforming $n-1$ independent, identically distributed (i.i.d.) $\Exp(1)$ random variables.
Finally, we note that we can recover these exact $n-1$ i.i.d. $\Exp(1)$ random variables
by applying the inverse transformation.

\subsubsection{Coalescent Posterior Predictive Check}
\label{subsec:cppc}

We consider the Bayesian approach for phylodynamic analysis laid out in
Section \ref{sec:methods} of the main text.
Similar to \citet{gelman1996posterior}'s mixed predictive distribution approach,
we simulate data and certain latent variables from our models,
informed by our posterior sample, in order to judge how well
those models adhere to observed and inferred realities.
In the context of our posterior with no sampling time model,
we replicate $\{\mathbf{y}^{\text{rep}}_i\}_{i=1}^{N}$
and $\{\mathbf{g}^{\text{rep}}_i\}_{i=1}^{N}$
according to this joint posterior,
\begin{equation}
\Pr(\mathbf{y}^{\text{rep}}, \mathbf{g}^{\text{rep}},
      \boldsymbol{\gamma}, \kappa, \boldsymbol{\theta} \mid
      \mathbf{y}, \mathbf{s}) \propto
      \Pr(\mathbf{y}^{\text{rep}} \mid \mathbf{g}^{\text{rep}}, \boldsymbol{\theta})
      \Pr(\mathbf{g}^{\text{rep}} \mid \boldsymbol{\gamma}, \mathbf{s})
      \Pr(\boldsymbol{\gamma}, \kappa, \boldsymbol{\theta} \mid \mathbf{y}, \mathbf{s}) ,
\label{eq:replicateposterior}
\end{equation}
simulating from the coalescent
$\Pr(\mathbf{g}^{\text{rep}} \mid \boldsymbol{\gamma}, \mathbf{s})$
and (if necessary, see below) the substitution model
$\Pr(\mathbf{y}^{\text{rep}} \mid \mathbf{g}^{\text{rep}}).$
We sample the final term on the right side via MCMC.

With posterior-sampled replicates available,
we construct a discrepancy $D_c$
\citep{gelman1996posterior, sinharay2003posterior}
on the observables and the inferred latent variables.
Let $G(\mathbf{g}, \boldsymbol{\gamma})$ be the transformation (explored in the previous section) that,
given the correct effective population trajectory,
and valid assumptions for the coalescent model,
will produce a sample of
$n-1$ i.i.d. $\Exp(1)$-distributed random variables.
Let $K$ be the Kolmogorov-Smirnov statistic \citep{massey1951kolmogorov},
\begin{equation}
K_{\Exp(1)}(\mathbf{e}) = \sup_{x \in \mathbb{R}} {|F_{\mathbf{e}}(x) - F_{\Exp(1)}(x)|},
\label{eq:K-Sstatistic}
\end{equation}
where $F_{\mathbf{e}}(x)$ is the
empirical cumulative distribution function (ECDF)
of $\mathbf{e}$,
and $F_{\Exp(1)}(x)$ is the
true cumulative distribution function (CDF)
of the $\Exp(1)$ distribution.
We define
\[
D_c(\mathbf{y}, \mathbf{g}, \mathbf{s}, \boldsymbol{\gamma}, \kappa) = 
      K_{\Exp(1)}(G(\mathbf{g}, \boldsymbol{\gamma})).
\]
Then when we run MCMC, we then compare the \textit{observed discrepancies},
\[\{D_c(\mathbf{y}, \mathbf{g}_i, \mathbf{s}, \boldsymbol{\gamma}_i, \kappa_i)\}_{i=1}^{N},\]
to the \textit{replicate discrepancies},
\[\{D_c(\mathbf{y}^{\text{rep}}_i, \mathbf{g}^{\text{rep}}_i, \mathbf{s}, \boldsymbol{\gamma}_i, \kappa_i)\}_{i=1}^{N}.\]
Note that the $D_c$ we constructed does not depend on $\mathbf{y}^{\text{rep}}$,
so we can save computation time by not simulating
$\mathbf{y}^{\text{rep}} \mid \mathbf{g}^{\text{rep}}$.
If we wish to check the sampling-aware posterior with the sampling time model,
the replicate posterior remains mostly the same as in Equation \ref{eq:replicateposterior},
but the final term becomes
${\Pr(\boldsymbol{\gamma}, \kappa, \boldsymbol{\beta}, \boldsymbol{\theta} \mid \mathbf{y}, \mathbf{s}, \mathcal{F})}$
to match the sampling-aware posterior. 

One method we have to compare the observed and replicate discrepancies is
the posterior predictive p-value \citep{gelman1996posterior}.
We calculate the posterior predictive p-value by finding the proportion of
MCMC iterations where the replicated discrepancy values are larger 
than its corresponding observed discrepancy value.
The smaller the posterior predictive p-value, the more unusual the observed data
is in the context of the chosen model.
Note that this posterior predictive p-value does not have the usual frequentist p-value
properties such as uniformity under a null model.
However, values close to 50\% suggest that the current model is adequate,
and for discrepancies that become larger as the observed data becomes less likely given a set of parameters,
the posterior predictive p-value tends to be smaller, to some degree, 
under under inadequate models \citep{gelman1996posterior}.

\subsubsection{Sampling Posterior Predictive Checks}
\label{subsec:sppc}

Similarly to the previous section,
we replicate $\{\mathbf{y}^{\text{rep}}_i\}_{i=1}^{N}$,
$\{\mathbf{g}^{\text{rep}}_i\}_{i=1}^{N}$,
and $\{\mathbf{s}^{\text{rep}}_i\}_{i=1}^{N}$
according to this joint posterior,
\begin{equation}
\begin{split}
\Pr(\mathbf{y}^{\text{rep}}, \mathbf{g}^{\text{rep}}, \mathbf{s}^{\text{rep}},
      \boldsymbol{\gamma}, \kappa, \boldsymbol{\beta}, \boldsymbol{\theta} \mid
      \mathbf{y}, \mathbf{s}) &\propto
      \Pr(\mathbf{y}^{\text{rep}} \mid \mathbf{g}^{\text{rep}})
      \Pr(\mathbf{g}^{\text{rep}} \mid \boldsymbol{\gamma}, \mathbf{s}^{\text{rep}})
      \Pr(\mathbf{s}^{\text{rep}} \mid \boldsymbol{\gamma}, \boldsymbol{\beta}) \\
      &\times {\Pr(\boldsymbol{\gamma}, \kappa, \boldsymbol{\beta}, \boldsymbol{\theta} \mid \mathbf{y}, \mathbf{s})},
\end{split}
\label{eq:Sreplicateposterior}
\end{equation}
with ${\Pr(\boldsymbol{\gamma}, \kappa, \boldsymbol{\beta}, \boldsymbol{\theta} \mid \mathbf{y}, \mathbf{s})}$
sampled via MCMC.
We simulate from the sampling model
${\Pr(\mathbf{s}^{\text{rep}} \mid \boldsymbol{\gamma}, \beta)}$,
and, if necessary, the coalescent
${\Pr(\mathbf{g}^{\text{rep}} \mid \boldsymbol{\gamma}, \mathbf{s})}$,
and the substitution model
${\Pr(\mathbf{y}^{\text{rep}} \mid \mathbf{g}^{\text{rep}})}.$


Suppose we divide the sampling interval into a grid $K_1, \ldots, K_l$,
potentially the same grid as used by grid-based priors for the effective population trajectory.
The sampling model is inhomogeneous Poisson,
so we can bin the numbers of sampling times
within each interval $m_1, \ldots, m_l$, each with expected values
$E_i = \int_{K_i}{\lambda_s(t)dt}$.
A common approach to problems with independent Poisson bins is a Chi-squared test
with statistic $\chi_s^2 = \sum_{i=1}^l \frac{(m_i - E_i)^2}{E_i}$ \citep{pearson1900x}.
We can then define a discrepancy
\begin{equation}
D_{\chi^2}(\mathbf{y}, \mathbf{g}, \mathbf{s}, \boldsymbol{\gamma}, \kappa) =
 \sum_{i=1}^l \frac{(m_i - E_i)^2}{E_i},
\label{eq:chisqdiscrepancy}
\end{equation}
for $m_i$ and $E_i$ derived from $\mathbf{s}$ as above.

\subsection{Results}

\subsubsection{Simulation Study}

\paragraph{Genealogy Inference}

We perform a simulation study in order to explore the capabilities of the posterior predictive checks
proposed above in Sections \ref{subsec:cppc} and \ref{subsec:sppc}.
We begin with a simplified version of the phylodynamic data-to-inference methodology.
Here we take genealogies to be our observed data
(and move on to inference based on observed sequence data in the next section).
We simulate sampling times according to inhomogeneous Poisson processes with 
different intensity trajectories via a time-transformation method \citep{Cinlar1975}
as we implemented in our R package \texttt{phylodyn} \citep{Karcher2017}.
Give sampling time data, we simulate from the coalescent model
using a similar time-tranformation method for the coalescent \citep{slatkin1991pairwise},
again as implemented in \texttt{phylodyn}.
For all of our fixed-tree simulations, we use an effective population size trajectory
designed to mimic the seasonal effective population size changes of a seasonal disease
such as influenza in North America \citep{zinder2014seasonality},
defined as follows:
\begin{equation}
\label{eq:ne_apb}
N_{e,l,u,p,o}(t) = \begin{cases} l + \frac{(u-l)}{1+\exp\{2[3-(\frac{t+o}{p} \, (\text{mod } 12))]\}}, & \mbox{if } \frac{t+o}{p}\, (\text{mod } 12) \leq 6, \\[10pt]
l + \frac{(u-l)}{1+\exp\{2[3 + (\frac{t+o}{p} \, (\text{mod } 12) - 12]\}} , & \mbox{if } \frac{t+o}{p} \, (\text{mod } 12) > 6. \end{cases}
\end{equation}
Specifically, we use $N_{e,10,100,12,0}(t)$ which is most comparable to 
an influenza effective population size trajectory as measured in units of weeks,
with $t=0$ representing the summer effective population size minimum.
We compare the results of our posterior predictive checks across different sampling scenario
and choice-of-posterior combinations.

\begin{figure}
	\centering
	\includegraphics[width=\textwidth]{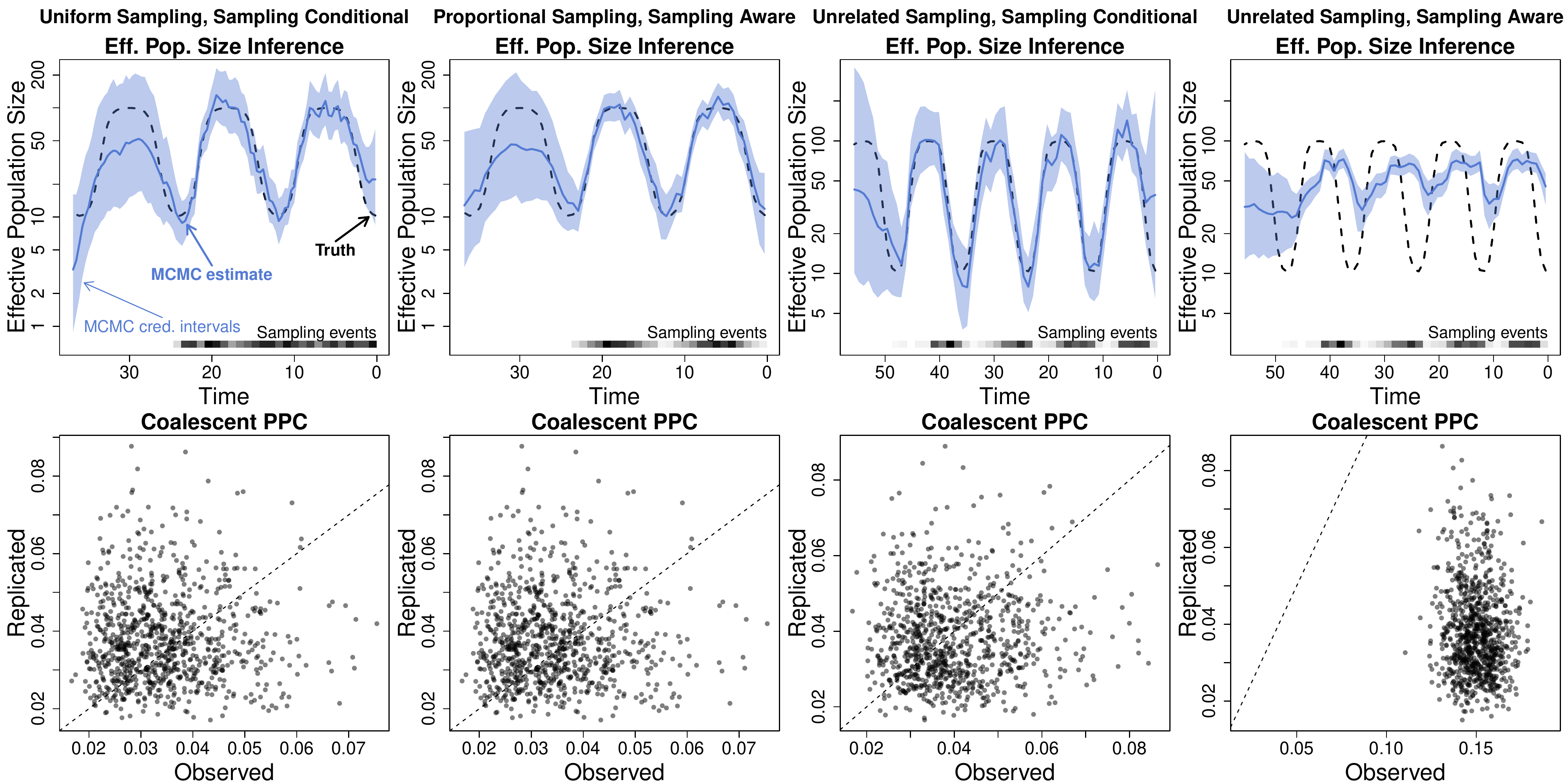}
	\caption{{\bf Effective population size inference and coalescent posterior predictive check for fixed-tree simulations.}
		The dashed black line represents the true effective population trajectory.
		The solid blue line represents the posterior median effective population trajectory
		inferred by fixed-tree MCMC 
		and the light blue region represents the corresponding pointwise 95\% credible intervals 
		for the effective population trajectory.}
	\label{fig:fixed_CPPC}
\end{figure}

\begin{figure}
	\centering
	\includegraphics[width=0.7\textwidth]{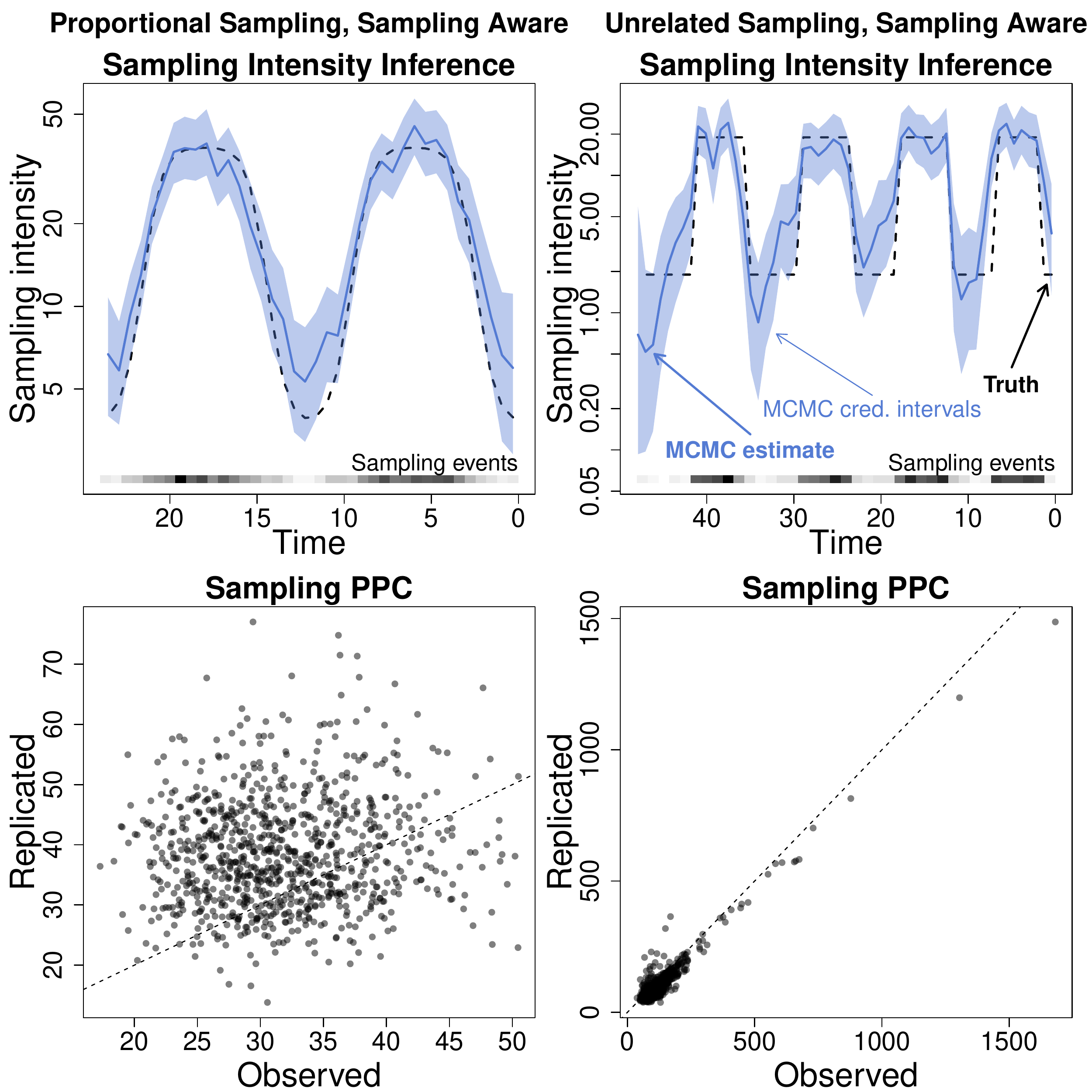}
	\caption{{\bf Sampling intensity inference and sampling time posterior predictive check for fixed-tree simulations.}
		The dashed black line represents the true sampling intensity.
		The solid blue line represents the posterior median sampling intensity
		inferred by fixed-tree MCMC,
		and the light blue region represents the corresponding pointwise 95\% credible intervals 
		for the sampling intensity.}
	\label{fig:fixed_SPPC}
\end{figure}

\begin{table}
\centering
\begin{tabular}{cccc}
\hline
 \multirow{2}{*}{\textbf{Scenario}} &  \multirow{2}{*}{\textbf{Sampling Model}} & \multicolumn{2}{c}{\textbf{Post. Pred. p-val}}\\
 & & Coalescent & Sampling \\
\hline
\rowcolor{Gray}
Uniform & Conditional & 0.58 & ---\\
Proportional & Aware: $\boldsymbol{\gamma}(t)$ & 0.59 & 0.72\\
\rowcolor{Gray}
Unrelated & Conditional & 0.46 & ---\\
Unrelated & Aware: $\boldsymbol{\gamma}(t)$ & 0.00 & 0.15\\
\hline
\end{tabular}
\caption{
{\bf Posterior predictive p-values for simulated fixed-tree data.}
}
\label{tab:fixed_tree_pvals}
\end{table}


In our first scenario, we simulate 500 sampling times, 
distributed according to a uniform distribution between $t=0$ and $t=24$ (weeks),
and simulate a genealogy with effective population size $N_{e,10,100,12,0}(t)$.
We infer the underlying effective population size trajectory 
with a sampling-conditional posterior 
using a Markov chain Monte Carlo (MCMC) method with 
an elliptical slice sampling transition kernel (ESS) \citep{murray2010elliptical}
as implemented in \texttt{phylodyn} 
(illustrated in the first row, first column of Figure \ref{fig:fixed_CPPC})
We use the MCMC output to generate replicate coalescent data as laid out 
in Section \ref{subsec:cppc} and calculate our coalescent discrepancy $D_c$
for the observed MCMC results as well as for the replicated results.
We plot the discrepancy comparison in the second row, first column of Figure \ref{fig:fixed_CPPC},
and note that the posterior predictive p-value is 0.58, which is close to 0.5, 
correctly suggesting that the model is adequate.

We proceed with several additional scenarios.
We simulate 514 sampling times between $t=0$ and $t=24$ (weeks), 
distributed proportionally to the effective population size,
with sampling log-intensity $\log[\lambda_c(t)] = -0.97 + N_{e,10,100,12,0}(t)$.
We infer the underlying effective population size trajectory
with a sampling-aware posterior 
(illustrated in the second column of Figure \ref{fig:fixed_CPPC}),
with sampling time model $\log[\lambda_s(t)] = \beta_0 + \beta_1 \cdot \boldsymbol{\gamma}(t)$.
We calculate the posterior predictive p-value as 0.59, again correctly suggesting adequacy.
We also simulate 509 sampling times between $t=0$ and $t=48$ (weeks),
distributed proportionally to a piecewise constant function $P(t)$ 
(illustrated in the second column of Figure \ref{fig:fixed_SPPC})
unrelated to the effective population size,
with log-sampling intensity $\log[\lambda_c(t)] = -1.67 + P(t)$.
We infer the underlying effective population size trajectory
using two different methods.
We use the sampling-conditional method 
(illustrated in the third column of Figure \ref{fig:fixed_CPPC})
and the sampling-aware method
(illustrated in the fourth column of Figure \ref{fig:fixed_CPPC})
with sampling log-intensity $\log[\lambda_s(t)] = \beta_0 + \beta_1 \cdot \boldsymbol{\gamma}(t)$.
The sampling-conditional posterior predictive p-value becomes 0.46,
suggesting that this method (which only considers the coalescent model)
does produce an adequate estimate of the effective population size trajectory.
The sampling-aware posterior predictive p-value becomes zero,
suggesting that this method produced a very poor estimate of the effective population size trajectory
(very visible in Figure \ref{fig:fixed_CPPC}).
This is likely due to the sampling time model mistaking fluctuations in sampling intensity
for information about the effective population size trajectory,
illustrating the importance of model checking when the true sampling model is uncertain.

For our sampling-aware scenarios,
we apply our sampling time posterior predictive check as well.
Our chi-squared sampling discrepancy $D_{\chi^2}$ generates a posterior predictive p-value of 0.72,
correctly suggesting a good fit.
The unrelated sampling scenario also produces a sampling posterior predictive p-value.
We see a relatively low posterior predictive p-value of 0.15,
reacting to differences between the true and inferred sampling intensity trajectories.

\paragraph{Sequence Data Inference}

\begin{figure}
	\centering
	\includegraphics[width=\textwidth]{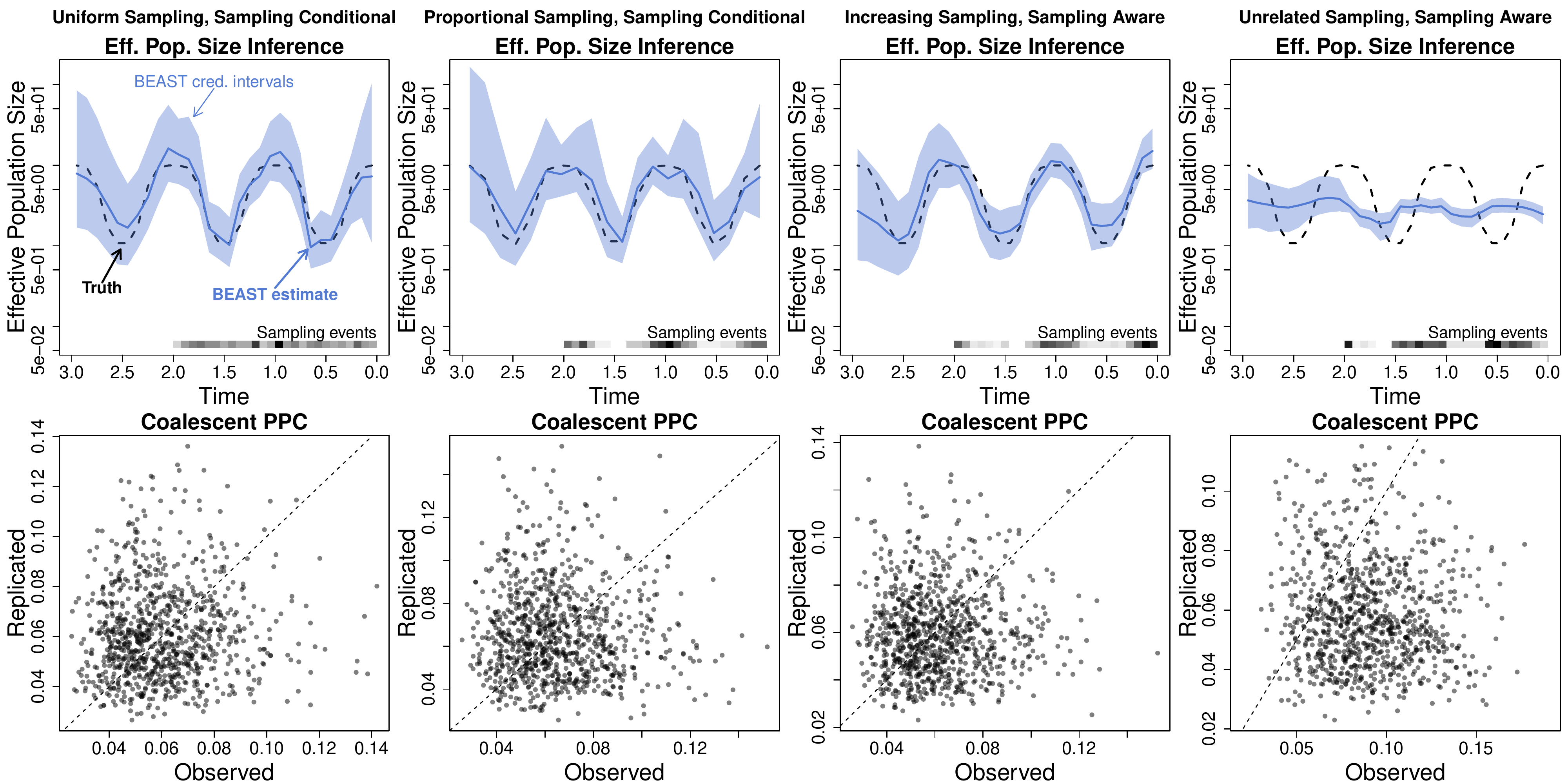}
	\caption{{\bf Effective population size inference and coalescent posterior predictive check for sequence data simulations.}
		The dashed black line represents the true effective population trajectory.
		The solid blue line represents the posterior median effective population trajectory
		inferred by BEAST,
		and the light blue region represents the corresponding pointwise 95\% credible intervals 
		for the effective population trajectory.}
	\label{fig:seqgen_CPPC}
\end{figure}

\begin{figure}
	\centering
	\includegraphics[width=0.7\textwidth]{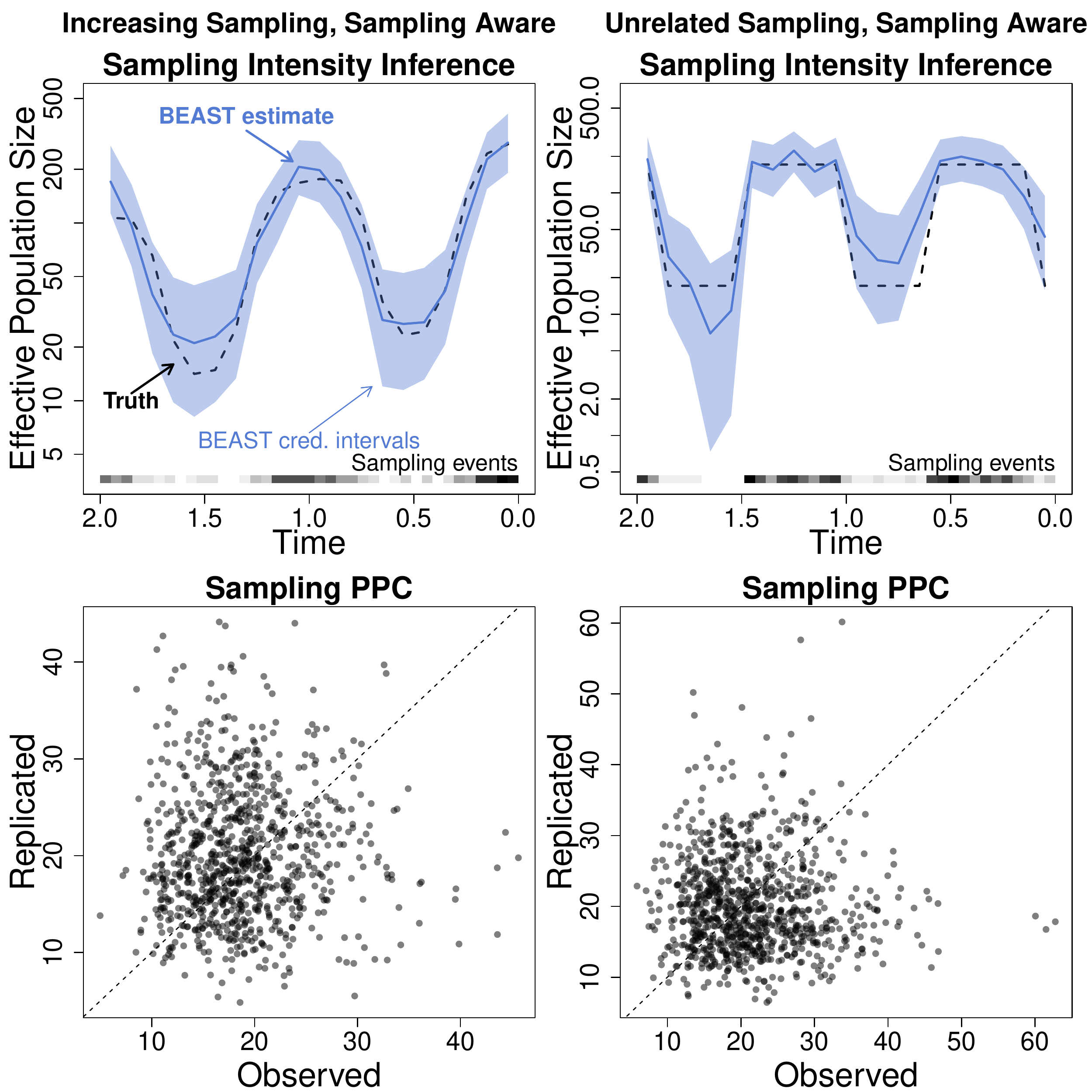}
	\caption{{\bf Sampling intensity inference and sampling time posterior predictive check for sequence data simulations.}
		The dashed black line represents the true sampling intensity.
		The solid blue line represents the posterior median sampling intensity
		inferred by BEAST,
		and the light blue region represents the corresponding pointwise 95\% credible intervals 
		for the sampling intensity.}
	\label{fig:seqgen_SPPC}
\end{figure}

\begin{table}
\centering
\begin{tabular}{cccc}
\hline
 \multirow{2}{*}{\textbf{Scenario}} &  \multirow{2}{*}{\textbf{Sampling Model}} & \multicolumn{2}{c}{\textbf{Post. Pred. p-val}}\\
 & & Coalescent & Sampling \\
\hline
\rowcolor{Gray}
Uniform & Conditional & 0.51 & ---\\
Proportional & Conditional & 0.50 & --- \\
\rowcolor{Gray}
Increasing & Aware: $\boldsymbol{\gamma}(t)$ & 0.47 & 0.56\\
Unrelated & Aware: $\boldsymbol{\gamma}(t)$ & 0.17 & 0.46\\
\hline
\end{tabular}
\caption{
{\bf Posterior predictive p-values for simulated sequence data.}
}
\label{tab:seqgen_pvals}
\end{table}


Now, we expand the scope of our simulation study to be based on 
simulated sequence alignment data instead of a known genealogy.
In this section, all of our examples will be based on an effective population size trajectory
of $N_{e,1,10,1,0.5}(t)$, mimicking the trajectory of a seasonal disease as measured in units of years.
Similar to the previous section, 
we generate sampling times and genealogies according to 
different sampling scenarios and the coalescent, respectively.
Given a genealogy, 
we simulate sequence data using the software \texttt{SeqGen} \citep{rambaut1997seq}
using the Jukes-Cantor 1969 \citep{jukes1969evolution} substitution model to generate 1500 sites.
We set the substitution rate to produce an expected 0.9 mutations per site, 
in order to produce a sequence alignment with many sites having one mutation 
and some sites having zero or multiple mutations.

For our first simulation, we distribute 200 sampling times uniformly between $t=0$ and $t=2$ (years).
We infer the underlying genealogy and effective population size trajectory using
the software \texttt{BEAST} \citep{BEAST} with an elliptical slice sampling transition kernel (ESS) \citep{murray2010elliptical}
as implemented in 
Section \ref{sec:methods}, 
with a sampling-conditional posterior. 
Finally, we generate replicate genealogies as in the previous section, 
and we calculate our coalescent discrepancy $D_c$ for the observed BEAST results
as well as the replicates.
In Figure \ref{fig:seqgen_CPPC} (first column), 
we see that the effective population estimate is close to the true trajectory, 
and when we compare the observed and replicate discrepancies,
we calculate a posterior predictive p-value of 0.51,
corroborating the model's adequacy.
Next, we distribute 170 sampling times between $t=0$ and $t=2$ (years) with sampling log-intensity
$\log[\lambda_c(t)] = 2.90 + N_{e,1,10,1,0.5}(t)$.
We infer the underlying genealogy and effective population size trajectory using
the sampling-conditional model and calculate discrepancies as above.
Note this is a model misspecification applying a sampling-conditional model 
to a preferential sampling sampling scenario in the style of \citep{karcher2015quantifying}.
Unfortunately, the posterior predictive p-value (0.50) does not detect this mismatch,
as the bias effective population size estimate is hard to visually detect in Figure \ref{fig:seqgen_CPPC}.

In our third scenario, we distribute 199 sampling times between $t=0$ and $t=2$ (years) 
with increasing sampling log-intensity $\log[\lambda_c(t)] = 3.35 - 0.5t + N_{e,1,10,1,0.5}(t)$.
We infer as above, but targeting the sampling-aware posterior 
with sampling log-intensity $\log[\lambda_s(t)] = \beta_0 + \beta_1 \cdot \boldsymbol{\gamma}(t)$.
This is again a misspecification, as the model cannot recover the $-0.5t$ term.
However, the posterior predictive check does not clearly detect the mismatch,
with a posterior predictive p-value of 0.47.
Our sampling posterior predictive check does not detect the misspecification either,
with a posterior predictive p-value of 0.56.
In our final scenario, we distribute 222 sampling times between $t=0$ and $t=2$ (years) 
with a sampling log-intensity $\log[\lambda_c(t)] = 2.84 + P'(t)$ 
($P'(t)$ illustrated in Figure \ref{fig:seqgen_SPPC}, second column) 
unrelated to the effective population size.
We target the sampling-aware posterior, 
with sampling log-intensity $\log[\lambda_s(t)] = \beta_0 + \beta_1 \cdot \boldsymbol{\gamma}(t)$.
The model reconstructs the effective population size trajectory poorly,
and this is successfully reflected in the posterior predictive p-value of 0.17.
However, our sampling posterior predictive check does not detect the misspecification,
with a posterior predictive p-value of 0.46.

\subsubsection{Seasonal Influenza}

\begin{figure}
	\centering
	\includegraphics[width=1.0\textwidth]{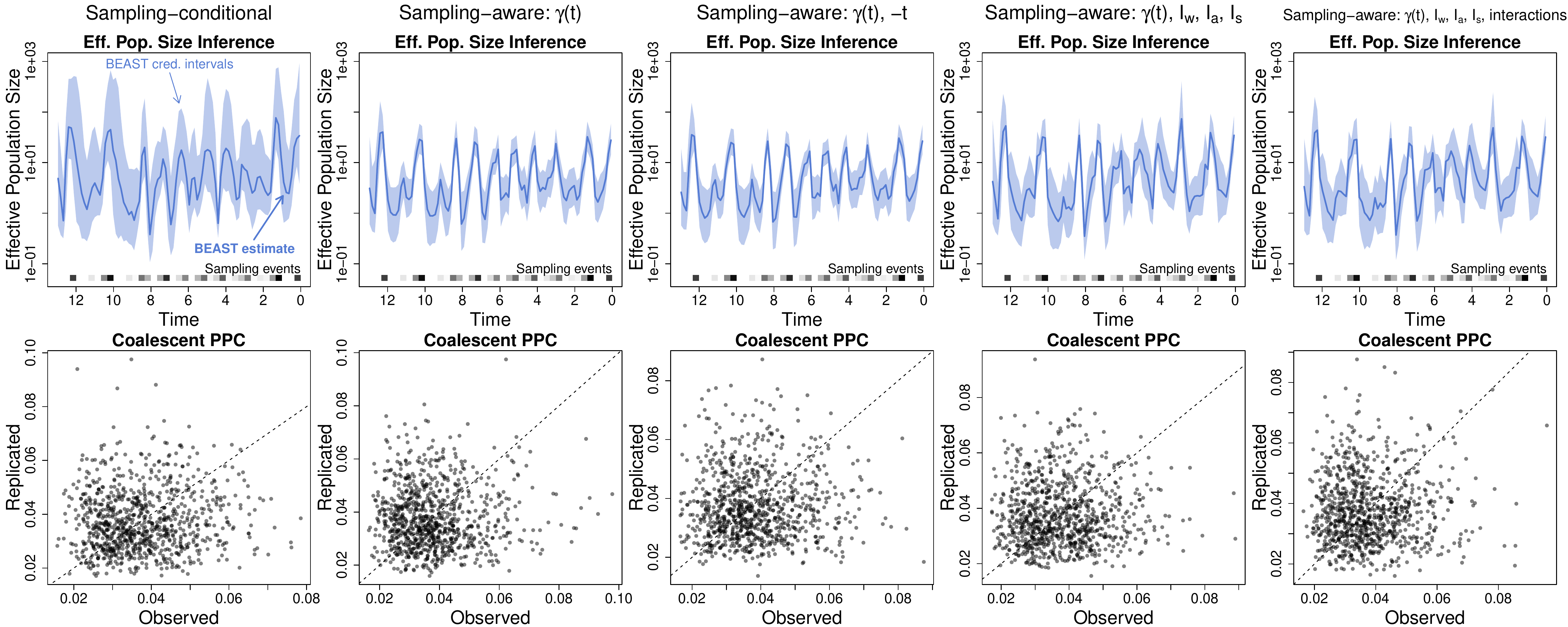}
	\caption{{\bf Effective population size inference and coalescent posterior predictive check for seasonal influenza data.}
		The solid blue line represents the posterior median effective population trajectory
		inferred by BEAST,
		and the light blue region represents the corresponding pointwise 95\% credible intervals 
		for the effective population trajectory.}
	\label{fig:USACanada_CPPC}
\end{figure}

\begin{figure}
	\centering
	\includegraphics[width=1.0\textwidth]{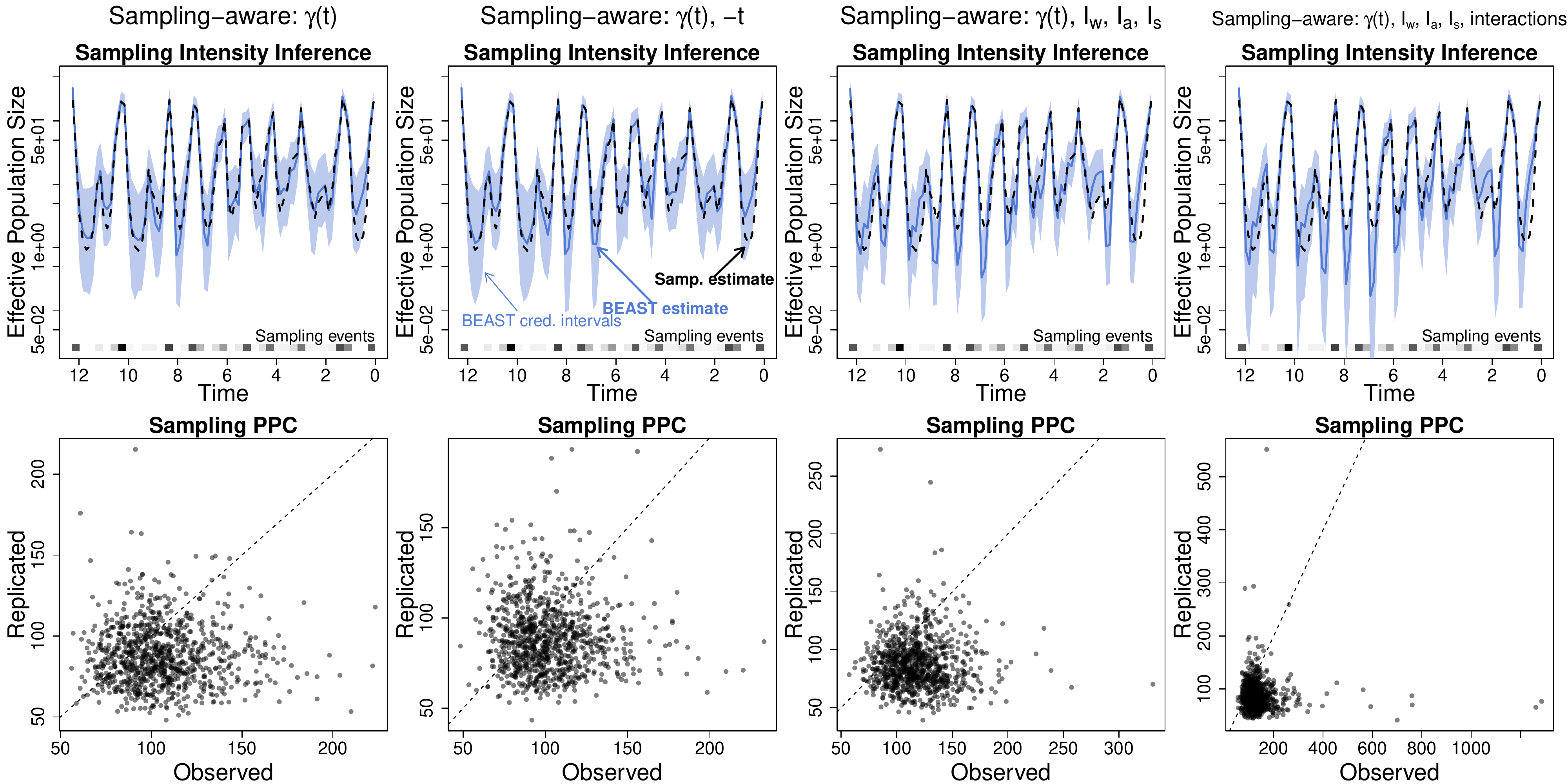}
	\caption{{\bf Sampling intensity inference and sampling time posterior predictive check for seasonal influenza data.}
		The dashed black line represents the true sampling intensity.
		The solid blue line represents the posterior median sampling intensity
		inferred by BEAST,
		and the light blue region represents the corresponding pointwise 95\% credible intervals 
		for the sampling intensity.}
	\label{fig:USACanada_SPPC}
\end{figure}

We apply our posterior predictive check methods to the North American subset of global H3N2 influenza \citep{zinder2014seasonality}.
The data contains 520 sequences aligned to form a multiple sequence alignment with 1698 sites of the hemagglutinin gene.
We use the same sequence data \texttt{BEAST} framework as the previous section,
choosing four different specific sampling time models.
We use a sampling-conditional model with no sampling time model,
a simple log-linear sampling time model,  
and sampling models with different sets of covariates,
including $I_w(t) = I_{(t \text{ mod } 1) \in [0,0.25)}$ as an indicator function for winter,
$I_a(t) = I_{(t \text{ mod } 1) \in [0.25,0.5)}$ as an indicator function for autumn,
and $I_s(t) = I_{(t \text{ mod } 1) \in [0.5,075)}$ as an indicator function for summer.

Figure \ref{fig:USACanada_CPPC} shows the inferred effective population size trajectories
and coalescent posterior predictive checks for the models.
All estimated trajectories follow a similar seasonal trajectory,
and the discrepancy comparison suggests that the estimated trajectory produces reasonable results
with large posterior predictive p-values (Table \ref{tab:USACanada_pvals}).
Figure \ref{fig:USACanada_SPPC} shows the inferred sampling intensities compared against 
a nonparametric sampling time-only estimate of the sampling intensity,
as well as sampling posterior predictive checks for the four models.
The sampling posterior predictive check produces moderate-to-low posterior predictive p-values,
suggesting some model inadequacy manifesting in the sampling intensity estimates.

\begin{table}
\centering
\begin{tabular}{ccc}
\hline
 \multirow{2}{*}{\textbf{Sampling Model}} & \multicolumn{2}{c}{\textbf{Post. Pred. p-val}}\\
 & Coalescent & Sampling \\
\hline
\rowcolor{Gray}
Conditional & 0.47 & ---\\
Aware: $\boldsymbol{\gamma}(t)$ & 0.48 & 0.29\\
\rowcolor{Gray}
Aware: $\boldsymbol{\gamma}(t), -t$ & 0.47 & 0.32\\
Aware: $\boldsymbol{\gamma}(t), I_w, I_a, I_s$ & 0.49 & 0.16\\
\rowcolor{Gray}
Aware: $\boldsymbol{\gamma}(t), I_w, I_a, I_s, \{I_w, I_a, I_s\} \cdot \boldsymbol{\gamma}(t) $ & 0.49 & 0.16\\
\hline
\end{tabular}
\caption{
{\bf Posterior predictive p-values for seasonal influenza data.}
}
\label{tab:USACanada_pvals}
\end{table}


\subsubsection{Ebola Outbreak}

Next, we analyze a subset of sequence data from the recent African Ebola outbreak \citep{dudas2017virus}.
We use the same sequence data \texttt{BEAST} framework as the previous section,
choosing four different specific sampling time models.
We use a sampling-conditional model with no sampling time model,
a simple log-linear sampling time model, 
and several additional sampling-aware models with different sets of covariates.

\begin{figure}
	\centering
	\includegraphics[width=\textwidth]{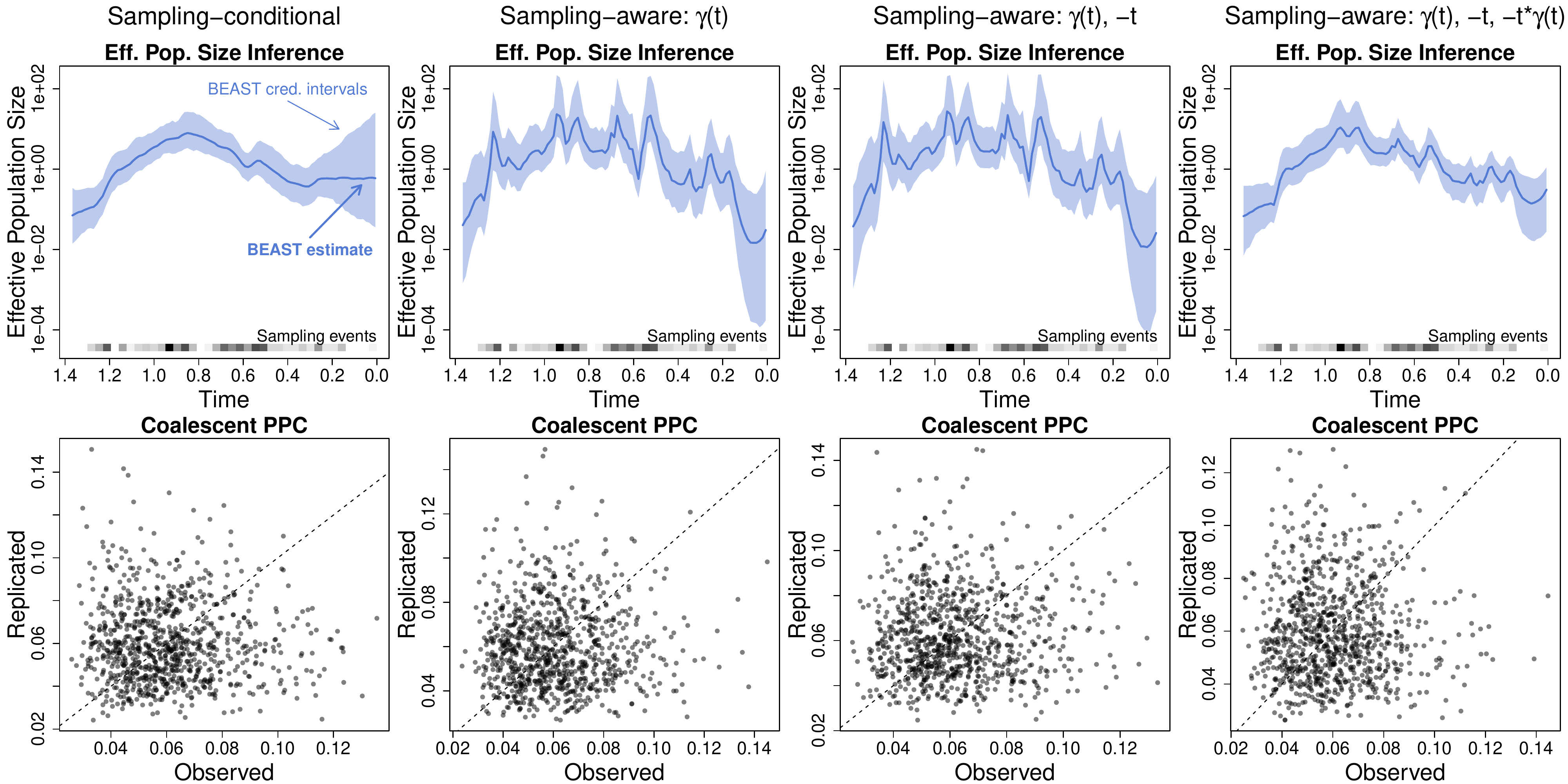}
	\caption{{\bf Effective population size inference and coalescent posterior predictive check for Sierra Leone Ebola data (part 1).}
		The solid blue line represents the posterior median effective population trajectory
		inferred by BEAST,
		and the light blue region represents the corresponding pointwise 95\% credible intervals 
		for the effective population trajectory.}
	\label{fig:Ebola_CPPC}
\end{figure}

\begin{figure}
	\centering
	\includegraphics[width=\textwidth]{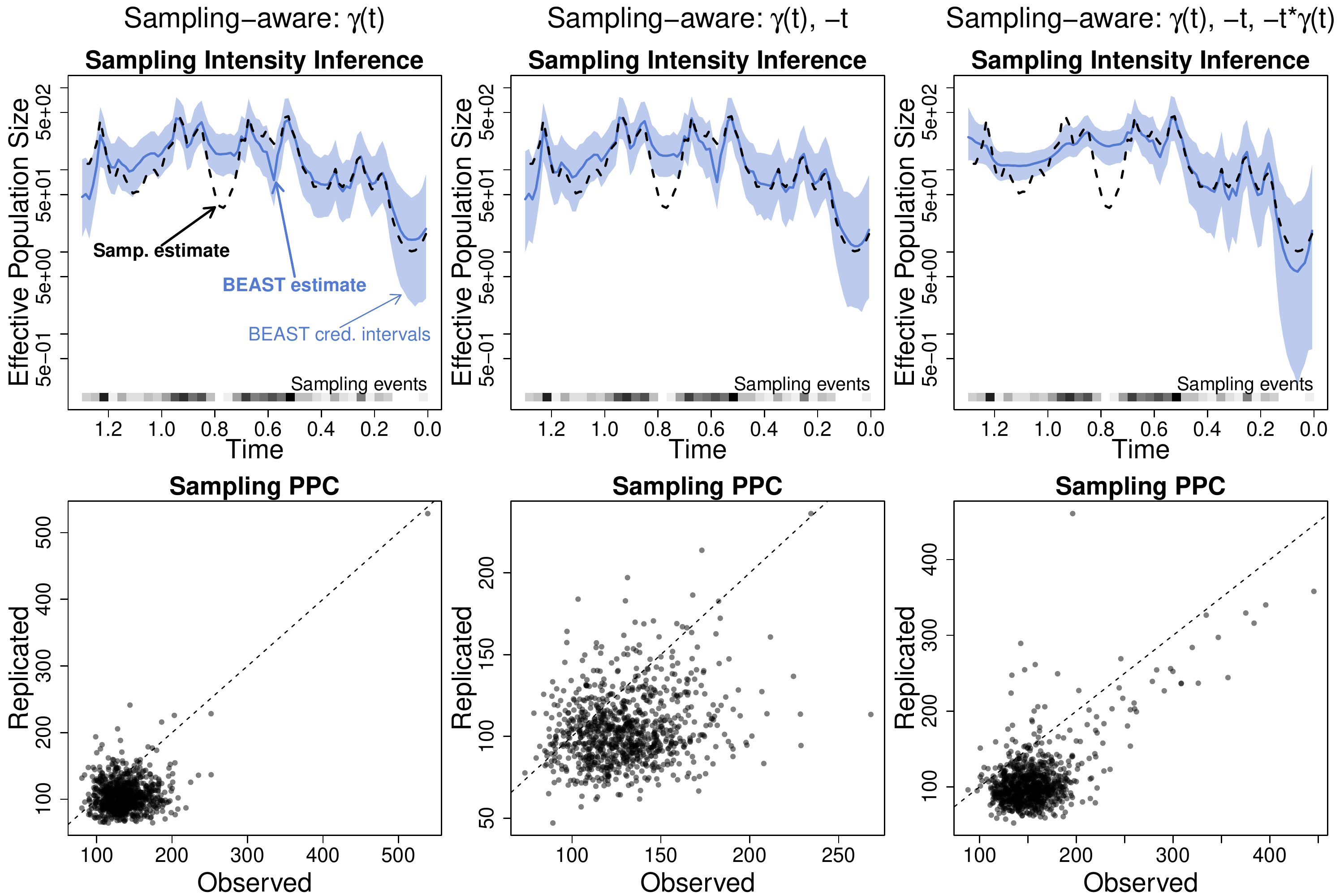}
	\caption{{\bf Sampling intensity inference and sampling time posterior predictive check for Sierra Leone Ebola data (part 1).}
		The dashed black line represents the true sampling intensity.
		The solid blue line represents the posterior median sampling intensity
		inferred by BEAST,
		and the light blue region represents the corresponding pointwise 95\% credible intervals 
		for the sampling intensity.}
	\label{fig:Ebola_SPPC}
\end{figure}

Figure \ref{fig:Ebola_CPPC} shows the inferred effective population size trajectories
and coalescent posterior predictive checks for the four models.
All estimated trajectories follow a similar effective population size trajectory 
that visually resembles a typical time trajectory of prevalence or incidence that peaks in Autumn of 2014.
The discrepancy comparison suggests that the estimated trajectory produces reasonable results
with large posterior predictive p-values (Table \ref{tab:Ebola_pvals}).
Figure \ref{fig:Ebola_SPPC} shows the inferred sampling intensities compared against 
a nonparametric sampling time-only estimate of the sampling intensity,
as well as sampling posterior predictive checks for the four models.
The sampling posterior predictive check produces small posterior predictive p-values (Table \ref{tab:Ebola_pvals}),
suggesting notable model inadequacy manifesting in the sampling intensity estimates.

\begin{figure}
	\centering
	\includegraphics[width=\textwidth]{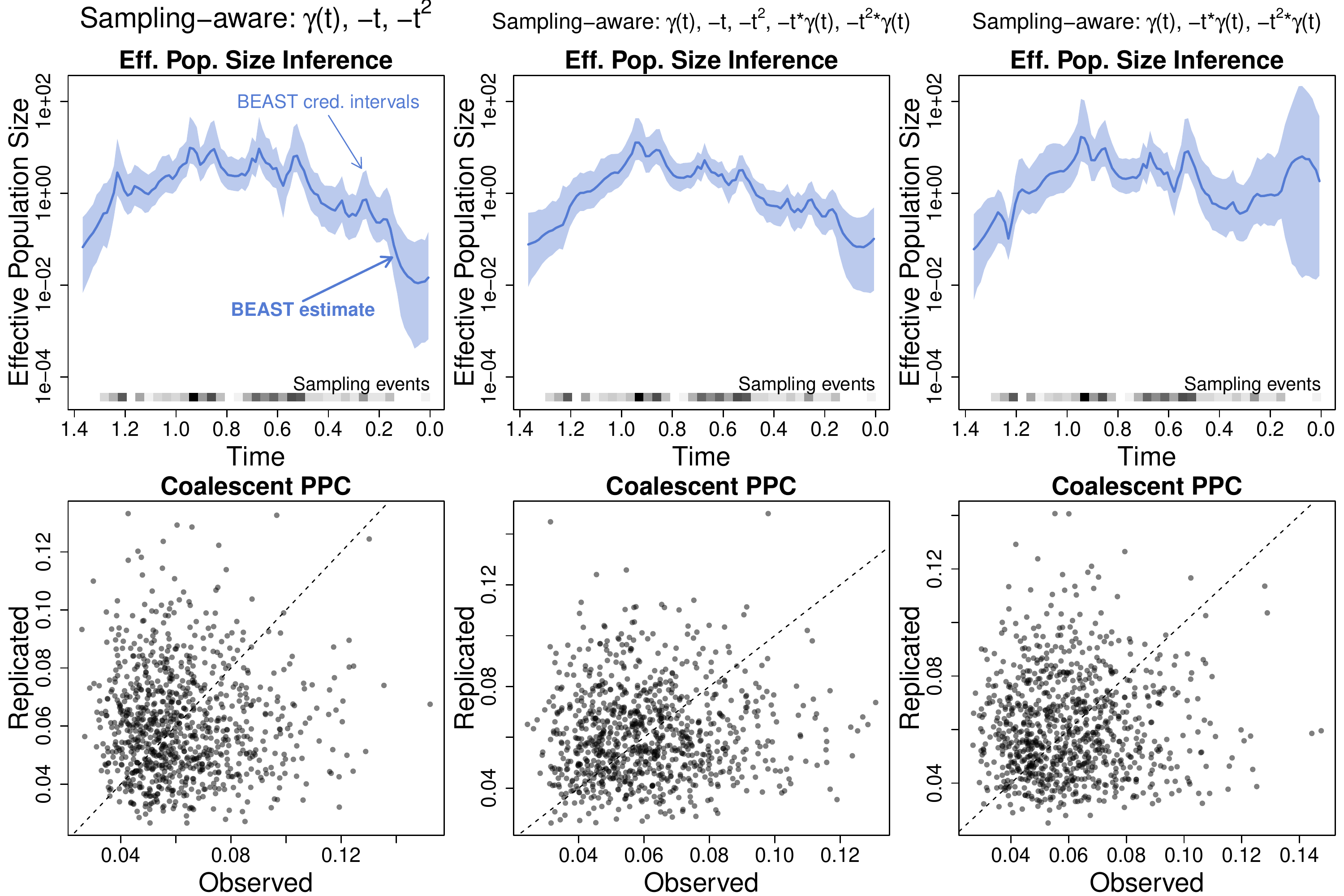}
	\caption{{\bf Effective population size inference and coalescent posterior predictive check for Sierra Leone Ebola data (part 2).}
		The solid blue line represents the posterior median effective population trajectory
		inferred by BEAST,
		and the light blue region represents the corresponding pointwise 95\% credible intervals 
		for the effective population trajectory.}
	\label{fig:Ebola_CPPC_supp}
\end{figure}

\begin{figure}
	\centering
	\includegraphics[width=\textwidth]{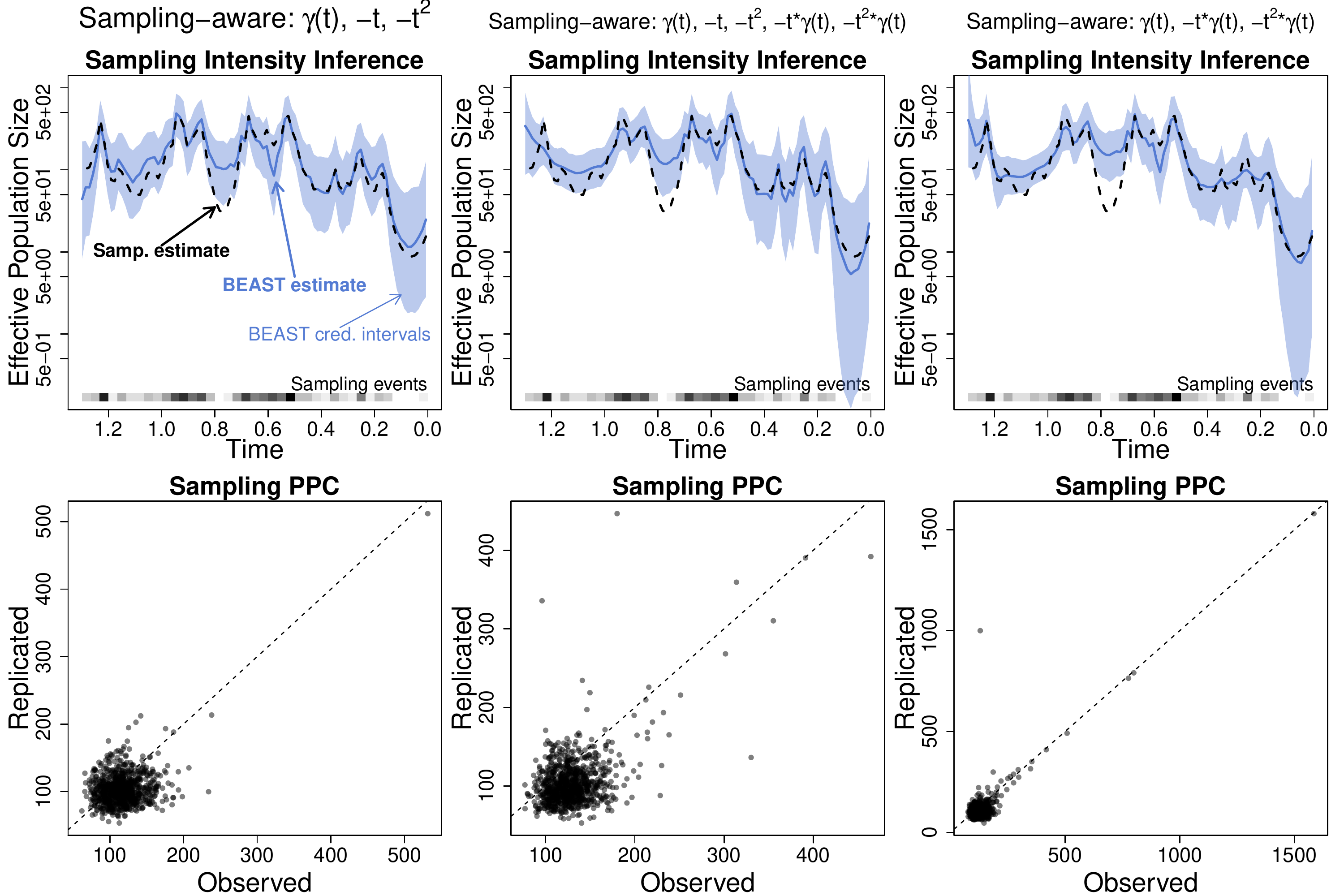}
	\caption{{\bf Sampling intensity inference and sampling time posterior predictive check for Sierra Leone Ebola data (part 2).}
		The dashed black line represents the true sampling intensity.
		The solid blue line represents the posterior median sampling intensity
		inferred by BEAST,
		and the light blue region represents the corresponding pointwise 95\% credible intervals 
		for the sampling intensity.}
	\label{fig:Ebola_SPPC_supp}
\end{figure}

\begin{table}
\centering
\begin{tabular}{lcc}
\hline
 \multirow{2}{*}{\textbf{Sampling Model}} & \multicolumn{2}{c}{\textbf{Post. Pred. p-val}}\\
 & Coalescent & Sampling \\
\hline
\rowcolor{Gray}
Conditional & 0.48 & ---\\
Aware: $\boldsymbol{\gamma}(t)$ & 0.47 & 0.15\\
\rowcolor{Gray}
Aware: $\boldsymbol{\gamma}(t), -t$ & 0.50 & 0.18\\
Aware: $\boldsymbol{\gamma}(t), -t, -t \cdot \boldsymbol{\gamma}(t)$ & 0.50 & 0.06\\
\rowcolor{Gray}
Aware: $\boldsymbol{\gamma}(t), -t, -t^2$ & 0.48 & 0.31\\
Aware: $\boldsymbol{\gamma}(t), -t, -t^2,  \{-t , -t^2\} \cdot \boldsymbol{\gamma}(t)$ & 0.51 & 0.17\\
\rowcolor{Gray}
Aware: $\boldsymbol{\gamma}(t), \{-t , -t^2\} \cdot \boldsymbol{\gamma}(t)$ & 0.51 & 0.22\\
\hline
\end{tabular}
\caption{
{\bf Posterior predictive p-values for Sierra Leone Ebola data.}
}
\label{tab:Ebola_pvals}
\end{table}

\clearpage
\pagenumbering{arabic}
\renewcommand*{\thepage}{C-\arabic{page}}

\renewcommand{\thefigure}{C-\arabic{figure}}
\setcounter{figure}{0}

\renewcommand{\thetable}{C-\arabic{table}}
\setcounter{table}{0}

\section{Appendix: Ebola incidence data}
\begin{figure}
	\centering
	\includegraphics[width=\textwidth]{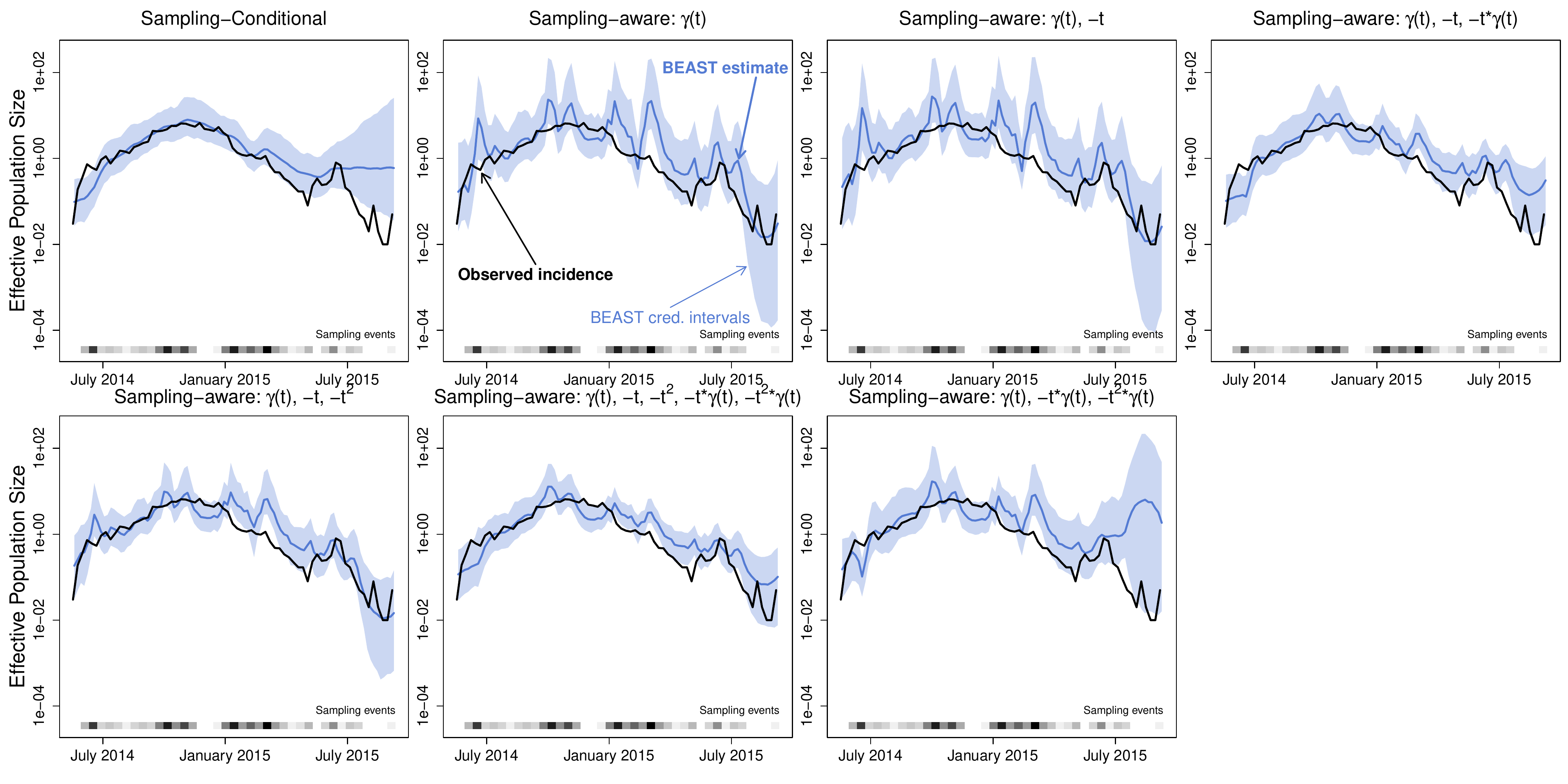}
	\caption{{\bf Comparison of effective population size reconstructions with incidence data in Sierra Leone.}
		The solid blue line represents the posterior median effective population trajectory
		inferred by BEAST,
		and the light blue region represents the corresponding pointwise 95\% credible intervals 
		for the effective population trajectory. The solid black line shows the observed incidence scaled by a constant so it is on the same scale as the effective population size. }
	\label{fig:Ebola_incidence_Sierra_Leone}
\end{figure}

\begin{figure}
	\centering
	\includegraphics[width=\textwidth]{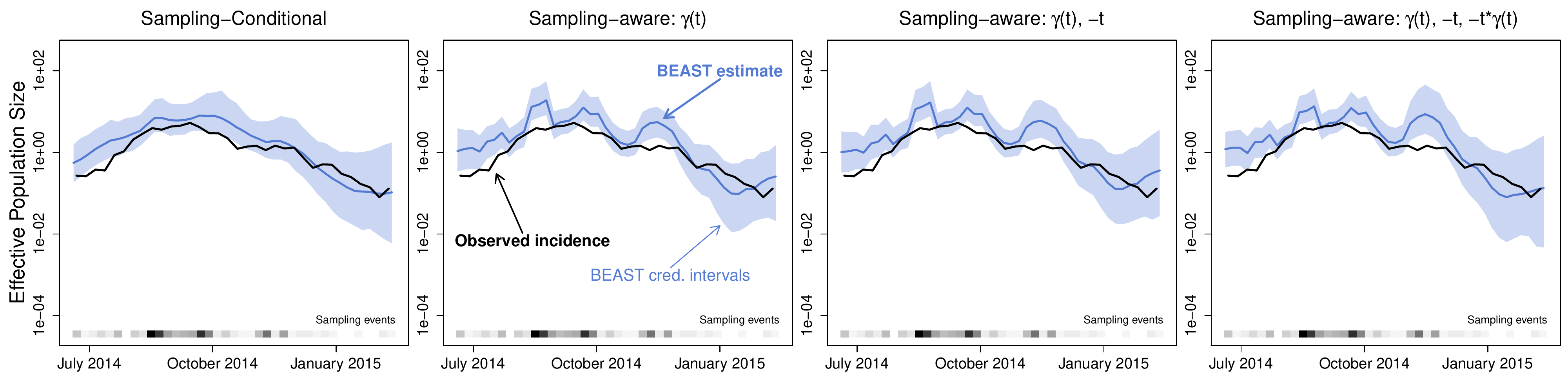}
	\caption{{\bf Comparison of effective population size reconstructions with incidence data in Liberia.} See Figure~\ref{fig:Ebola_incidence_Sierra_Leone} caption for the legend explanation.}
	\label{fig:Ebola_incidence_Liberia}
\end{figure}

As we explain in Section~\ref{ebola_results}, we compare estimated effective population size trajectories with observed incidence data. 
We multiplied incidence count by 0.01 --- a number determined by trial-and-error, to bring incidence and effective population size to the same scale --- and plot effective population size posterior summaries and incidence counts for Sierra Leone and Liberia in Figures \ref{fig:Ebola_incidence_Sierra_Leone} and \ref{fig:Ebola_incidence_Liberia}.

\end{document}